\def\half{\frac{1}{2}}
\def\hi#1#2{$#1$\kern -2pt-#2} 
\def\hy#1#2{#1-\kern -2pt$#2$} 

\documentstyle[aps,graphicx,subfigure]{revtex}

\begin{document}

\title{{\Large \bf Phase Shifts and Resonances in the Dirac Equation}}
\author{{\large \bf Piers Kennedy $^{*a}$ , Richard L. Hall $^{+b}$ and Norman Dombey $^{\dagger a}$ }}
\address{ ${^a}$Centre for Theoretical Physics, University of Sussex, Brighton BN1 9QJ, UK\\
${^b}$Department of Mathematics and Statistics, Concordia University, Montreal, Quebec, H3G 1M8, Canada\\
$^{*}$email: kapv4@sussex.ac.uk ~ $^{+}$email: rhall@mathstat.concordia.ca ~ $^{\dagger}$email: normand@sussex.ac.uk }

\hspace*{4.85in}CUQM-103\\
\hspace*{5in}math-ph/0401015\\
\hspace*{5in}January 2004
\vspace{0.2in}
\begin{center}
{\Large \bf Phase Shifts and Resonances in the Dirac Equation}\\
\vspace{0.1in}
{\large \bf Piers Kennedy $^{*a}$, Richard L. Hall $^{+b}$, and Norman Dombey $^{\dagger a}$ }\\
\end{center}
\vspace{0.2in}
\noindent\hspace*{0.5in}
${^a}$Centre for Theoretical Physics, University of Sussex, Brighton BN1 9QJ, UK\\
\noindent\hspace*{0.5in} ${^b}$Department of Mathematics and Statistics, Concordia University, Montreal, Quebec, H3G 1M8, Canada\\
\noindent\hspace*{0.5in} $^{*}$email: kapv4@sussex.ac.uk ~ $^{+}$email: rhall@mathstat.concordia.ca ~ $^{\dagger}$email: normand@sussex.ac.uk 

\section*{Abstract}
We review the analytic results for the phase shifts $\delta _{l}(k)$ in
non-relativistic scattering from a spherical well. The conditions for the
existence of resonances are established in terms of time-delays. Resonances
are shown to exist for p-waves (and higher angular momenta) but {\sl not}
for s-waves. These resonances occur when the potential is not quite strong
enough to support a bound p-wave of zero energy. We then examine
relativistic scattering by spherical wells and barriers in the Dirac
equation. In contrast to the non-relativistic situation, s-waves are now
seen to possess resonances in scattering from both wells {\sl and} barriers.
When s-wave resonances occur for scattering from a well, the potential is not
quite strong enough to support a zero momentum s-wave solution at $E=m$. Resonances resulting
from scattering from a barrier can be explained in terms of the `crossing'
theorem linking s-wave scattering from barriers to p-wave  scattering from
wells. A numerical procedure to extract phase shifts for general short range
potentials is introduced and illustrated by considering relativistic
scattering from a Gaussian potential well and barrier.

\section*{Introduction} 

The question of what does or does not constitute a resonance in potential
scattering still seems to cause some confusion. Many authors who consider
scattering in simple systems such as the one-dimensional square well in the
Schr\"{o}dinger equation talk of transmission `resonances' when the
transmission coefficient $T=1$ with little or no justification as to why
these might be resonances. It would seem on the surface that a resonance
exists  on account of  a maximum or peak in a measure of the scattering (for
example the transmission coefficient in one dimension or the cross-section
for a particular partial wave in three dimensions). But Wigner \cite{Wigner}
pointed out nearly 50 years ago that in order to have a true physical
resonance the scattered particle must be captured in some way by the
scattering centre. So in addition to the peak in the scattering probability
function, there must also be a delay in the transition time of the particle
through the target: this requirement is fulfilled by demanding that the
Wigner time delay is positive. This implies  that the phase shift
associated with the peak must {\sl increase} through an odd multiple of $%
\frac{\pi }{2}$ as the particle energy increases through the resonance
energy.

We will begin this paper by looking at the phase shifts for the simple case
of scattering from a spherical well \footnote{this is often referred to as the 
three-dimensional square well} in the Schr\"{o}dinger equation in
order to illustrate resonant behaviour and distinguish it from cases where
the cross-section peaks but there is no resonance. The argument and results
closely follow those given by Newton \cite{Newton}. It will be seen that,
using Wigner's criteria, the spherical well does not give rise to s-wave
resonances. Resonances do  exist, however,  for scattered p-waves and waves
of higher angular momenta. This comes about because of  the centrifugal term
in the Schr\"{o}dinger equation for partial waves $l>0$. 

The relationship between bound states and continuum states for
non-relativistic systems is well documented. Schiff \cite{Schiff} noted that 
{`}a potential well that has an energy level nearly at zero exhibits a
resonance in the low-energy scattering of particles with the same $l$ value
as the energy level'. Newton \cite{Newton} later studied this more carefully
and showed that one can go smoothly from a zero energy solution to a real resonance 
by weakening the potential slightly. This idea
is crucial to the the procedure we will adopt. First the condition on the
potential to support a zero energy solution must be established. This
condition can then be relaxed slightly to push the solution into a
continuum state. Finally the phase shift corresponding to the resulting 
peak in cross-section  is checked to see if it fulfils the criterion of
positive time delay necessary for a real resonance.

In this paper we generalise these results to the relativistic Dirac equation
in three dimensions. We begin by reproducing the results for the phase shift
in the spherical potential well and analyse whether resonant
behaviour exists.  In contrast to the non-relativistic results it will be
shown that scattered s-waves do now give rise to resonances. It will be
shown that these arise because of the presence of a centrifugal term in the
original coupled equations which does not disappear even for s-waves. But we
cannot stop there: in the Dirac equation we must consider the zero momentum solutions 
which exist at both $E=m$ and $E=-m$. We refer to these as critical solutions and 
in particular it is conventional to refer to the solution at $E=-m$ as supercritical. 
It will be shown that a `crossing theorem' exists which
connects scattering of particles by potential barriers to the supercritical
state at $E=-m.$ This phenomenon was first illustrated in a previous paper
by two of us \cite{DH} in the context of positron-heavy nuclei scattering. We use
this theorem to show that resonances exist in s-wave Dirac scattering from
spherical barriers as well as spherical wells. Resonances also occur in higher
partial waves as expected.

We then investigate resonant scattering by general short range spherically-symmetric 
potential wells and barriers in the Dirac equation. A numerical procedure for the
extraction of the phase shifts is demonstrated in the case of a Gaussian potential
wells and barriers, and the results are analysed. We demonstrate that s-wave
resonances can exist in scattering by both barriers and wells.

\section*{The Schr\"odinger equation}

The phase shifts $\delta _{l}(k)$  for the spherical well  $V(r)=-V$%
\quad $r\leq a;$\quad $V(r)=0$  $r>a$ are well known \cite{Schiff}, 
\cite{J}: 
\begin{equation}
\tan \delta _{l}=\frac{k\,j_{l}^{\prime
}(ka)j_{l}(pa)-p\,j_{l}(ka)j_{l}^{\prime }(pa)}{k\,n_{l}^{\prime
}(ka)j_{l}(pa)-p\,n_{l}(ka)j_{l}^{\prime }(pa)}  
\label{res1}
\end{equation}
where $j_{l}$ and $n_{l}$ are the regular and irregular spherical Bessel
functions respectively, $k^{2}=2mE$ and $p^{2}=2m(E+V)$. So for s-wave
scattering when $l=0$ this becomes: 
\begin{equation}
\tan \delta _{0}=\frac{k\tan pa-p\tan ka}{p+k\tan ka\tan pa}  
\label{res2}
\end{equation}
while for p-wave scattering when $l=1$ this gives: 
\begin{equation}
\tan \delta _{1}=\frac{akp^{2}\tan pa-ak^{2}p\tan ka+(k^{2}-p^{2})\tan
ka\tan pa}{ak^{2}p+(p^{2}-k^{2})\tan pa+akp^{2}\tan ka\tan pa}  
\label{res3}
\end{equation}
The potential wells of the greatest interest are those which are not quite
deep enough to support a bound state. The critical values of the potential
well which support states of zero energy (these may or may not be bound -
this will be discussed later) are well established \cite{Schiff}. For $l=0$
we have 
\begin{equation}
j_{0}(pa)=pa\,j_{1}(pa) \\
\Rightarrow \quad \cos \,pa=0\quad \Rightarrow V_{c}=\frac{(2n-1)^{2}\pi ^{2}%
}{8ma^{2}}  
\label{res4}
\end{equation}
and for $l>0$ we have 
\begin{equation}
j_{l-1}(pa)=0  
\label{res5}
\end{equation}
For $l=1$ this leads to 
\begin{equation}
\sin \,pa=0\quad \Rightarrow \quad V_{c}=\frac{n^{2}\pi ^{2}}{2ma^{2}}
\label{res6}
\end{equation}
We are now in a position to examine the low energy phase shift behaviour in
order to see if resonances occur. In the case of the s-wave, if the
potential is not quite strong enough to bind the first state, the
phase shift $\delta _{0}(k)$ will be seen to rise from zero but never
quite reach the value $\frac{\pi }{2}$ before it decreases for larger
energies. If we continue to increase the strength of the potential well
there will be a value at which the phase shift at zero energy $\delta
_{0}(0)$ flips from zero to $\frac{\pi }{2}$. The strength at this point is
exactly the critical value $V=V_{c}$ from Eq. (\ref{res4}), required for
the potential to just support the first $l=0$ critical state. As soon as we
just increase the strength of the potential from $V_c$, $%
\delta _{0}(0)$ flips to the value $\pi $ and decreases for increasing
energy. This behaviour is illustrated in the following figure: 
\begin{figure}[tbph]
\par
\begin{center}
\leavevmode
\includegraphics[width=0.4\linewidth]{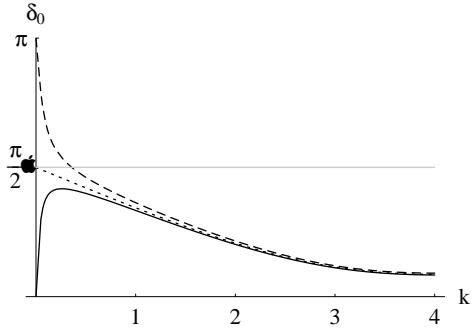} \medskip
\caption{The s-wave phase shifts for the spherical well. A zero-energy state exists for a potential of
strength $V_{c}=0.25\pi ^{2}\approx 2.47$ (dotted line). The solid
line represents a potential of strength $V=2.4$ which almost binds whilst
the dashed line is the potential of strength $V=2.6$ which contains a
loosely bound state. ($a=1$ and $m=0.5$).}
\end{center}
\label{Sswps}
\end{figure}

The process repeats with every additional bound state, so $\delta _{0}(0)$
increases by $\pi $ each time and we have Levinson's theorem: 
\begin{equation}
\delta _{0}(0)=n_{0}\pi  
\label{res7}
\end{equation}
where $n_{0}$ tells us the number of $l=0$ bound states. For critical values
of the potential when it is just strong enough to possess a zero-energy
 state, Levinson's theorem must be modified to 
\begin{equation}
\delta _{0}(0)=(n_{0}+\displaystyle \half)\pi  
\label{res8}
\end{equation}
Although it is not obvious, the critical state is not actually bound but
instead forms part of the continuum. In order to see this consider the
Schr\"{o}dinger equation with for a potential well: 
\begin{equation}
\Psi ^{\prime \prime }(r)-\left[ \frac{l(l+1)}{r^{2}} + 2mV(r)-k^{2}\right]
\Psi (r)=0  
\label{res9}
\end{equation}
If the potential vanishes faster than $1/r^{2}$ at large distances, the
zero-energy solution for large $r$ is found from 
\begin{equation}
\Psi ^{\prime \prime }(r)\approx \frac{l(l+1)}{r^{2}}\Psi (r)  
\label{res10}
\end{equation}
Consequently $\Psi $ behaves as $1/r^{l}$ at infinity and the particle will
only be normalisable, hence bound, for $l>0$. This means that for the first
critical value of the potential, $V=V_{c}=\pi ^{2}/4$ when $l=0$, the state is not
bound, $n_{0}=0$ and $\delta _{0}(0)=\frac{\pi }{2}$ from (8). These states
are often referred to as half-bound states \cite{Newton}.

In order for a resonance to exist the Wigner time delay must be positive.
This is defined as 
\begin{equation}
\tau_{l} =2\frac{d\delta_{l} }{dE}=\frac{2}{v}\frac{d\delta_{l} }{dk}
\label{res11}
\end{equation}
So the phase shift must increase through $\pi /2$. A decrease in the
phase shift as the energy increases through $\pi /2$ gives rise to a
negative time delay - a time advancement \cite{Kamke}. This does {\sl not}
constitute a resonance. Hence there are no s-wave resonances for a spherical
well.

The results for p-waves (and other higher angular momentum waves) are
distinctly different. If the potential is not quite strong enough to bind
the first state, the phase shift $\delta _{1}(k)$ rises from zero but it
now increases through the value $\frac{\pi }{2}$ at some positive value of
the energy up to a maximum value ($<\pi $) before decreasing for larger
energies. If we continue to increase the strength of the potential, the
maximum of the phase shift approaches the value $\pi $ for decreasing values
of the energy until the phase shift at zero energy $\delta _{1}(0)$ flips
from zero to $\pi $. The strength of the potential at this point is exactly
the critical value $V_{c}=\pi ^{2}$ from (6), required for the potential to
just support the first $l=1$ bound state. It should be noted that, contrary
to the s-wave case, the critical wave functions are bound states.

To illustrate that there are p-wave resonances but no s-wave resonances, we
can consider the effective potential $V_{eff}(r)=V(r)+l(l+1)/r^{2}$ shown in
figure 2. 
\begin{figure}[tbph]
\par
\begin{center}
\leavevmode
\includegraphics[width=0.7\linewidth]{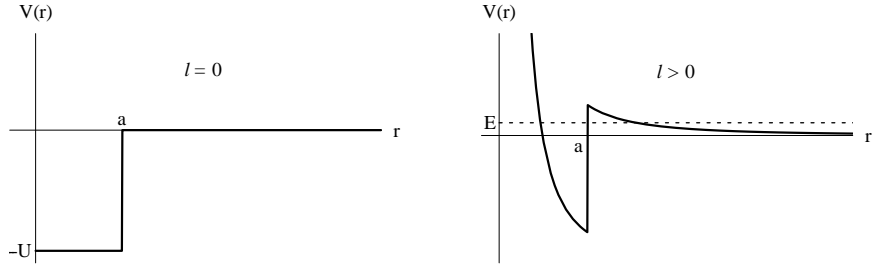} \medskip
\caption{The effective potential for a spherical well.}
\end{center}
\label{effpot}
\end{figure}
For $l=0$ there is no centrifugal barrier to trap the particle for small
positive energies and consequently low momentum resonances do not occur. At
zero energy there is nothing to stop the particle escaping the well - it is
not bound and forms part of the continuum. For angular momenta $l>0$, there
is now a barrier which can momentarily trap particles giving rise to a
resonance in the physical region $E>0$. If we scatter a particle of energy $%
E<l(l+1)/a^{2}$ (i.e. below the lip of the repulsive centrifugal barrier -
see figure 2), it can tunnel into the target and once there eventually leak
back out. If we deepen the potential the particle becomes increasingly
trapped as the strength of the barrier through which it must tunnel
increases and the resonance becomes more pronounced as it is retarded for
greater lengths of time. We finally arrive at the critical value of the
potential for which $E=0$ and the particle is completely trapped provided $%
l>0.$

As we continue to increase the strength of the potential beyond this
critical value $\delta _{l}(0)$ jumps by $\pi $ for each additional bound
state as is the case for $l=0$. For $l>0$ there are no half-bound states so
Eq. (8) is no longer applicable and Eq. (7) can be modified to 
\begin{equation}
\delta _{l}(0)=n_{l}\pi  
\label{res12}
\end{equation}
where $n_{l}$ now tells us the number of bound states with angular momentum $%
l$. From figure (3) we can see that the phase shift associated with
potentials which almost contain a bound state will increase through an odd
multiple of $\frac{\pi }{2}$. As we get closer to criticality
the phase shift increases more and more rapidly through this value and will
peak at values that tend to $0 $ (mod $\pi $). This manifests itself as an
increasingly more pronounced peak in the cross-section.

This behaviour is illustrated in figure 3. The solid line represents a potential of strength $V=9$ and the dashed line represents a potential of strength $V=9.7$, where both potentials are just too weak to have a bound state. The dotted line is a potential of strength $V=10$ which contains a loosely bound state. A zero-energy $l=1$ bound state exists for a potential well of strength $V_c=\pi^2 \approx 9.87$. ($a=1$ and $m=0.5$). 

\begin{figure}[tbph]
\par
\begin{center}
\leavevmode
\includegraphics[width=0.4\linewidth]{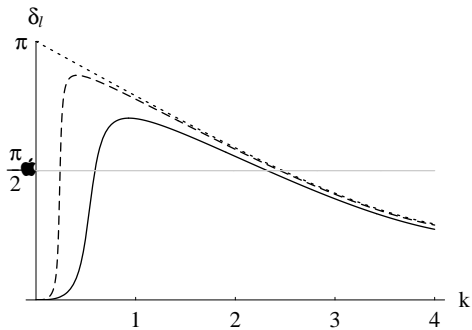} \medskip
\caption{The spherical well for p-wave phase shifts.}
\end{center}
\label{Spwps}
\end{figure}

\section*{The Dirac Equation : Scattering Solutions}

We will now consider the problem of scattering from a spherical well in the relativistic Dirac equation. Following Greiner, M\"uller and Rafelski \cite{GMR}, which we refer to as GMR, the coupled radial equations for an electron of mass $m$ and energy $E$ in the presence the spherical well $V(r)=-V$\quad $r\leq a;$\quad $V(r)=0$\quad  $r>a$ can be written
\begin{eqnarray}
f^{\prime}(r)&=& -\frac{\chi}{r}f(r)+(E+m+V)g(r)\nonumber \\
g^{\prime}(r)&=&-(E-m+V)f(r)+\frac{\chi}{r}g(r)
\label{res13}
\end{eqnarray}
where the Dirac wave function $\Psi(r)$ for the radial equation is given by $r\Psi(r)=\left(
\begin{array}{c}
f \\
g
\end{array}\right)$
The variable $\chi = \pm (j+\half)=\pm 1, \pm 2, ...$ with the orbital angular momentum 
$l=j+\half\frac{\chi}{|\chi|}=$\{$0,1,2,..$\} corresponding to \{$s,p,d,..$\} waves. When $r<a$, the solutions are
\begin{eqnarray}
f(r)&=&a_1 r j_{l_\chi}(pr)\nonumber \\
g(r)&=&a_1\frac{\chi}{|\chi|} \frac{pr}{E+V+m} j_{l_{-\chi}}(pr)
\label{res14}
\end{eqnarray}
where the well momentum $p$ is
\begin{equation}
p^2=(E+V)^2-m^2
\label{res15}
\end{equation}
and
\begin{equation}
l_{\chi} =\left\{\begin{array}{lr}
\chi &\qquad \chi>0\\
-\chi-1 &\qquad \chi<0
\end{array} \right. \qquad , \qquad
l_{-\chi} =\left\{\begin{array}{lr}
-\chi &\qquad -\chi>0\\
\chi-1 &\qquad -\chi<0
\end{array} \right. 
\label{res16}
\end{equation}
When $r>a$, the free particle solutions are:
\begin{eqnarray}
f(r)&=&b_1 r j_{l_\chi}(kr)+b_2 r n_{l_\chi}(kr)\nonumber \\
g(r)&=&\frac{\chi}{|\chi|} \frac{kr}{E+m}[b_1 j_{l_{-\chi}}(kr)+b_2 n_{l_{-\chi}}(kr)]
\label{res17}
\end{eqnarray}
where $k^2=E^2-m^2$. When $r \to \infty$ the following asymptotic forms for the spherical Bessel functions are required:
\begin{eqnarray}
x \, j_n(x) \sim & \sin \left( x- \frac{n \pi}{2} \right) \qquad\qquad
x \, n_n(x) \sim & -\cos \left( x- \frac{n \pi}{2} \right) 
\label{res18}
\end{eqnarray}
The components can therefore be written as
\begin{eqnarray}
f(r) &\to& \frac{1}{k} \left[ b_1 \sin \left( kr - \frac{l_\chi \pi}{2} \right) -b_2 \cos \left( kr - \frac{l_\chi \pi}{2} \right) \right]
\label{res19}\\
g(r) &\to& \frac{\chi}{|\chi|} \frac{1}{E+m} \left[ b_1 \sin \left( kr - \frac{l_{-\chi} \pi}{2} \right) -b_2 \cos \left( kr - \frac{l_{-\chi} \pi}{2} \right) \right] 
\label{res20}
\end{eqnarray}
Therefore 
\begin{eqnarray}
f(r) &\to& \frac{b_1}{|b_1|}\frac{A}{k}\sin \left( kr - \frac{l_\chi \pi}{2} + \delta_{l_\chi} \right) 
\label{res21}
 \\
g(r) &\to& \frac{b_1}{|b_1|} \frac{\chi}{|\chi|}\frac{A}{E+m} \sin \left( kr - \frac{l_{-\chi} \pi}{2}+ \delta_{l_\chi} \right)
\label{res22}
\end{eqnarray}
where
\begin{equation}
A \cos \delta_{l_\chi} = b_1 \qquad\qquad A \sin \delta_{l_\chi} = -b_2
\label{res23}
\end{equation}
So 
\begin{equation}
\tan \delta_{l_\chi} = -\frac{b_2}{b_1} \qquad\qquad A^2=b_1^2+b_2^2
\label{res24}
\end{equation}
This relationship between the phase shifts and the coefficients $b_1$ and $b_2$ was shown for s-waves by GMR. Depending on the sign of $\chi$ we have
\begin{eqnarray}
l_{-\chi} =\left\{\begin{array}{l}
l_\chi+1 \qquad \chi<0\\
l_{\chi}-1 \qquad \chi>0
\end{array} \right. \qquad\qquad
l_\chi-l_{-\chi}=\frac{\chi}{|\chi|}
\label{res25}
\end{eqnarray}
Using the above together with Eqs. (\ref{res21} and \ref{res22}), the two components are found to be
\begin{eqnarray}
f(r) &\to& \frac{b_1}{|b_1|}\frac{A}{k}\sin \left( kr - \frac{l_\chi \pi}{2} + \delta_{l_\chi} \right) 
\label{res26}\\
g(r) &\to& \frac{b_1}{|b_1|} \frac{A}{E+m} \cos \left( kr - \frac{l_{\chi} \pi}{2}+ \delta_{l_\chi} \right)
\label{res27}
\end{eqnarray}
The asymptotic forms are clearly dependent on the sign of $\chi$:
\begin{equation}
r\Psi(r) =\left(
\begin{array}{c}
f(r)\\
g(r)
\end{array}
\right)
\to \frac{b_1}{|b_1|} A\left( \begin{array}{c}
\frac{\displaystyle 1}{\displaystyle k}\cos \left( kr + \frac{\chi \pi}{2} + \delta_{l_\chi} \right)  \\[0.2cm]
- \frac{\displaystyle 1}{\displaystyle E+m} \sin \left( kr + \frac{\chi \pi}{2}+ \delta_{l_\chi} \right)
\end{array} \right)
\qquad\qquad \chi<0
\label{res28}
\end{equation}

\begin{equation}
r\Psi(r) =\left(
\begin{array}{c}
f(r)\\
g(r)
\end{array}
\right)
\to \frac{b_1}{|b_1|} A\left( \begin{array}{c}
\frac{\displaystyle 1}{\displaystyle k} \sin \left( kr - \frac{\chi \pi}{2} + \delta_{l_\chi} \right)  \\[0.2cm]
\frac{\displaystyle 1}{\displaystyle E+m} \cos \left( kr - \frac{\chi \pi}{2}+ \delta_{l_\chi} \right)
\end{array} \right)
\qquad\qquad \chi>0
\label{res29}
\end{equation}
These results are consistent with Barth\'el\'emy \cite{MCB1}, \cite{MCB2}. We can combine Eqs. (\ref{res17}) and (\ref{res24}) to incorporate the phase shift into the free-particle wave components of the wave function:
\begin{eqnarray}
f(r)&=&b_1 r[j_{l_\chi}(kr)-\tan \delta_{l_\chi} n_{l_\chi}(kr)]\nonumber \\
g(r)&=&\frac{\displaystyle \chi}{\displaystyle |\chi|} \frac{\displaystyle kr}{\displaystyle E+m}b_1 [j_{l_{-\chi}}(kr)-\tan \delta_{l_\chi} n_{l_{-\chi}}(kr)]
\label{res30}
\end{eqnarray}
The solutions for $r<a$ from Eqs. (\ref{res13}) and $r>a$ from Eqs. (\ref{res30}) must be joined smoothly; by dividing the top component by the bottom component and equating the two solutions at $r=a$ we arrive at
\begin{eqnarray}
\frac{a_1 j_{l_\chi}(pa)}{\frac{\chi}{|\chi|} \frac{p}{E+V+m}a_1 j_{l_{-\chi}}(pa)} = 
\frac{b_1 [j_{l_\chi}(ka)-\tan \delta_{l_\chi} n_{l_\chi}(ka)]}{\frac{\chi}{|\chi|} \frac{k}{E+m}b_1 [j_{l_{-\chi}}(ka)-\tan \delta_{l_\chi} n_{l_{-\chi}}(ka)]}
\label{res31}
\end{eqnarray}
After re-arrangement this becomes
\begin{equation}
\frac{k}{p}\, \frac{E+V+m}{E+m} \, \frac{j_{l_\chi}(pa)}{j_{l_{-\chi}}(pa)}= 
\frac{j_{l_\chi}(ka)-\tan \delta_{l_\chi} n_{l_\chi}(ka)}{j_{l_{-\chi}}(ka)-\tan \delta_{l_\chi} n_{l_{-\chi}}(ka)}
\label{res32}
\end{equation}
Taking $p=\sqrt{(E+V)^2-m^2}$ we find that the kinematic term $\gamma$ is given by
\begin{equation}
\gamma=\frac{p}{k}\, \frac{E+m}{E+V+m} 
\label{res33}
\end{equation}
Solving for $\tan \delta_{l_\chi}$ we find that the relativistic equivalent of the Schr\"odinger result (\ref{res1}) is
\begin{equation}
\tan \delta_{l_\chi} = \frac{\gamma j_{l_\chi}(ka)j_{l_{-\chi}}(pa)-j_{l_\chi}(pa)j_{l_{-\chi}}(ka)}{\gamma j_{l_{-\chi}}(pa)n_{l_\chi}(ka)-j_{l_\chi}(pa)n_{l_{-\chi}}(ka)}
\label{res34}
\end{equation}


\section*{The Dirac Equation : Zero Momentum Solutions}
In this section we establish the critical solutions for the spherical well. Unlike the previous non-relativistic situation, the Dirac equation has two critical solutions corresponding to $E=m$ and $E=-m$; the analysis which follows allows us to consider both of these critical solutions. For the spherical well potential the coupled radial Eqs. (\ref{res13}) have the solutions inside the well:
\begin{eqnarray}
f(r)&=&r[a_1 j_{l_{\chi}}(pr)+a_2n_{l_{\chi}}(pr)] \nonumber\\
g(r)&=&\frac{\chi}{|\chi|}\frac{pr}{E+V+m}[a_1 j_{l_{-\chi}}(pr)+a_2n_{l_{-\chi}}(pr)]
\label{res35}
\end{eqnarray}
and the solutions outside the well are
\begin{eqnarray}
f(r)&=&r\sqrt{\frac{2\kappa r}{\pi}}[b_1 K_{l_{\chi}+\half}(\kappa r)+b_2I_{l_{\chi}+\half}(\kappa r)] \nonumber\\
g(r)&=&\frac{\kappa r}{E+m}\sqrt{\frac{2\kappa r}{\pi}}[-b_1 K_{l_{-\chi}+\half}(\kappa r)+b_2I_{l_{-\chi}+\half}(\kappa r)]
\label{res36}
\end{eqnarray}
where $\kappa^2=m^2-E^2$. $K_{n+\half}(\kappa r)$ and $I_{n+\half}(\kappa r)$ are the modified spherical Bessel functions. 
In order to establish the bound states the internal wave function must be normalisable at the origin therefore $a_2=0$. The external wave function must also be normalisable at infinity which imposes the condition $b_2=0$. By matching the top and bottom components at $r=a$ the following conditions are obtained:
\begin{eqnarray}
a_1 j_{l_{\chi}}(pa) &=& \sqrt{\frac{2\kappa a}{\pi}}b_1 K_{l_{\chi}+\half}(\kappa a) \nonumber\\
\frac{\chi}{|\chi|} \frac{p}{E+V+m}a_1 j_{l_{-\chi}}(pa) &=& -\frac{\kappa}{E+m}\sqrt{\frac{2\kappa a}{\pi}}b_1 K_{l_{-\chi}+\half}(\kappa a)
\label{res37}
\end{eqnarray}
Dividing the first equation by the second gives:
\begin{equation}
\frac{\chi}{|\chi|}(E+V+m ) \frac{j_{l_{\chi}}(pa)}{ p \, j_{l_{-\chi}}(pa)}=
-(E+m)\frac{ K_{l_{\chi}+\half}(\kappa a)} {\kappa \, K_{l_{-\chi}+\half}(\kappa a)}
\label{res38}
\end{equation}
Upon re-arrangement this becomes
\begin{equation}
\frac{j_{l_{\chi}}(pa)}{j_{l_{-\chi}}(pa)}=-\frac{\chi}{|\chi|} \, \frac{p}{\kappa}\,
\frac{E+m}{E+V+m}\frac{ K_{l_{\chi}+\half}(\kappa a)} {K_{l_{-\chi}+\half}(\kappa a)}
\label{res39}
\end{equation}
The critical limits $E \to \pm m$ imply $\kappa \to 0$, which together with the following identity from reference \cite{Arfken}
\begin{equation}
\lim_{z \to 0} \left[ \sqrt{\frac{2}{\pi z}}K_{n+\half}(z)\right] \to (2n-1)!!z^{-n-1},
\label{res40}
\end{equation}
allow us to find the zero momentum limit of Eq. (\ref{res39}). Using Eq. (\ref{res15}) with $E=\pm m$ it is easy to see that $p \to p_{\pm} =\sqrt{V(V \pm 2m)}$. Eq. (\ref{res39}) becomes
\begin{eqnarray}
\frac{j_{l_{\chi}}(p_{\pm}a)}{j_{l_{-\chi}}(p_{\pm}a)}
\to \left. -\frac{\chi}{|\chi|} \, \frac{p_{\pm}}{\kappa}\frac{E+m}{E+V+m}\frac{(2l_\chi-1)!!}{(2l_{-\chi}-1)!!}(\kappa a)^{-(l_\chi-l_{-\chi})}
\right|_{E \to \pm m , \, \kappa \to 0} 
\label{res41}
\end{eqnarray}
We now consider $\chi<0$ and $\chi>0$ separately.
\subsection*{Case I : $\chi<0$}
When $\chi <0$ this simplifies to
\begin{eqnarray}
\frac{j_{-\chi-1}(p_{\pm}a)}{j_{-\chi}(p_{\pm}a)}
\to \left. -  \frac{p_{\pm}a}{1+2\chi}\frac{E+m}{E+V+m}
\right|_{E \to \pm m , \, \kappa \to 0} 
\label{res42}
\end{eqnarray}
For critical states at $E=m$, the above can be written as
\begin{eqnarray}
\frac{j_{-\chi-1}(p_{+}a)}{j_{-\chi}(p_{+}a)}=
-\frac{2ma}{1+2\chi}\sqrt{\frac{V}{V+2m}}
\label{res43}
\end{eqnarray}
The supercritical states at $E=-m$ satisfy
\begin{equation}
j_{-\chi-1}(p_{-}a)=0
\label{res44}
\end{equation}
In particular for critical s-waves ($\chi=-1$, $l_{\chi}=0$, $l_{-\chi}=1$) at $E=m$, Eq. (\ref{res43}) is
\begin{equation}
\frac{j_{0}(p_{+}a)}{j_{1}(p_{+}a)}= 2ma\sqrt{\frac{V}{V+2m}}
\label{res45}
\end{equation}
Upon simplification this reduces to the transcendental equation
\begin{equation}
\tan a \sqrt{V(V+2m)} =  -2ma\sqrt{\frac{V+2m}{V}}
\label{res46}
\end{equation}
Supercritical s-waves  satisfy $j_0(p_{-}a)=0\Rightarrow \sin(p_{-}a)=0$ from Eq. (\ref{res44}). This imposes the following condition on the well depth:
\begin{equation}
a\sqrt{V(V-2m)}=n\pi
\label{res47}
\end{equation}
\subsection*{Case II : $\chi>0$}
When $\chi>0$, Eq. (\ref{res40}) simplifies to
\begin{eqnarray}
\frac{j_{\chi}(p_{\pm}a)}{j_{\chi-1}(p_{\pm}a)}
=\left. -  \frac{p_{\pm}(2\chi-1)}{a}\frac{1}{(m-E)(E+V+m)}
\right|_{E \to \pm m , \, \kappa \to 0} 
\label{res48}
\end{eqnarray}
The critical solutions $E = m$ give
\begin{equation}
j_{\chi-1}(p_{+}a)=0
\label{res49}
\end{equation}
while the supercritical solutions at $E=-m$ are 
\begin{eqnarray}
\frac{j_{\chi}(p_{-}a)}{j_{\chi-1}(p_{-}a)}=
-\frac{2\chi-1}{2ma}\sqrt{\frac{V-2m}{V}}
\label{res50}
\end{eqnarray}
For critical p-waves ($\chi=1$, $l_{\chi}=1$, $l_{-\chi}=0$) at $E=m$, Eq. (\ref{res49}) imposes the condition $j_0(p_+ a)=0 \Rightarrow \sin(p_+a)=0$. The potential depth must therefore satisfy
\begin{equation}
a\sqrt{V(V+2m)}=n\pi
\label{res51}
\end{equation}
p-waves become supercritical when
\begin{equation}
\frac{j_{1}(p_{-}a)}{j_{0}(p_{-}a)} = -\frac{1}{2ma}\sqrt{\frac{V-2m}{V}}
\label{res52}
\end{equation}
Upon simplification this becomes
\begin{equation}
\tan a \sqrt{V(V-2m)} = 2ma \sqrt{\frac{V-2m}{V}}
\label{res53}
\end{equation}
These conditions can also be derived from the bound state eigenvalue equations provided in GMR.


\section*{Phase Shift Behaviour : The Spherical Well}

In this section we start by considering the conditions on the potential for a resonance to exist when low-momentum electrons are scattered on a spherical well. These conditions will then be shown to be related to the results for the critical solutions derived in the previous section. Finally the s-wave phase shift is explicitly derived and both s- and p-wave phase shifts are illustrated for spherical wells which nearly support an s or p zero momentum solution at $E=m$.

A necessary condition for the existence of a resonance is that the phase shift is $\delta_{l_{\chi}}=\frac{\pi}{2}$. Further analysis is required to establish if this position corresponds to an increasing phase shift to give a resonance. In order to ensure this the denominator on the right hand side of Eq. (\ref{res34}) must be zero which gives the condition
\begin{equation}
\frac{j_{l_{\chi}}(pa)}{j_{l_{-\chi}}(pa)}=
\gamma \frac{n_{l_{\chi}}(ka)}{n_{l_{-\chi}}(ka)}
\label{res54}
\end{equation}
For low momentum scattering $E \to m$ on the spherical well $p \to p_{+}$ and
\begin{equation}
\lim_{x \to 0}[x^{l+1}n_{l}(x)] \to -(2l-1)!!
\label{res55}
\end{equation}
Eq. (\ref{res54}) becomes
\begin{equation}
\frac{j_{l_{\chi}}(p_{+}a)}{j_{l_{-\chi}}(p_{+}a)}=
ka\gamma \frac{(2l_{\chi}-1)!!}{(2l_{-\chi}-1)!!}(ka)^{-(l_{\chi}-l_{-\chi}+1)}
\label{res56}
\end{equation}
Using Eq. (\ref{res33})
\begin{equation}
ka \gamma=ap_{+}\frac{E+m}{E+V+m} \to 2ma\sqrt{\frac{V}{V+2m}}
\label{res57}
\end{equation}
Eq. (\ref{res56}) becomes
\begin{equation}
\frac{j_{l_{\chi}}(p_{+}a)}{j_{l_{-\chi}}(p_{+}a)}=
2ma \sqrt{\frac{V}{V+2m}}\, \frac{(2l_{\chi}-1)!!}{(2l_{-\chi}-1)!!}(ka)^{-(l_{\chi}-l_{-\chi}+1)}
\label{res58}
\end{equation}
When $\chi <0$, this is
\begin{equation}
\frac{j_{-\chi-1}(p_{+}a)}{j_{-\chi}(p_{+}a)}=
-\frac{2ma}{1+2\chi} \sqrt{\frac{V}{V+2m}}
\label{res59}
\end{equation}
which is identical to Eq. (\ref{res43}), the condition for $\chi<0$ critical states to be at $E=m$. When $\chi >0$, the right hand side of Eq. (\ref{res58}) is infinite leading to the condition
\begin{equation}
j_{\chi-1}(p_{+}a)=0
\label{res60}
\end{equation}
which is identical to Eq. (\ref{res49}), the condition for $\chi>0$ critical states to be at $E=m$. The implications of these results will now be highlighted for s- and p-wave scattering.

For s-waves Eq. (\ref{res34}) simplifies to
\begin{equation}
\tan \delta_0  = \frac{\gamma j_0(ka)j_1(pa)-j_0(pa)j_1(ka)}{\gamma j_1(pa)n_0(ka)-j_0(pa)n_1(ka)}
\label{res61}
\end{equation}
Using standard identities for $j_l(x)$ and $n_l(x)$ with $l=0,1$ we find:
\begin{eqnarray}
\tan \delta_0 
= \frac{(\gamma k -1)\tan ka \tan pa - \gamma k p a \tan ka-ka \tan pa}{akp \tan ka \tan pa+(1-\gamma k)\tan pa+\gamma k pa}
\label{res62}
\end{eqnarray}
If we compare this to the non-relativistic case it can be seen that this more closely resembles the p-wave phase shift (Eq. \ref{res3}) than the corresponding s-wave phase shift (Eq. \ref{res2}). The difference in behaviour of this s-wave phase shift is clear in the figure below:
\begin{figure}[tbph]
\par
\begin{center}
\leavevmode
\includegraphics[width=0.4\linewidth]{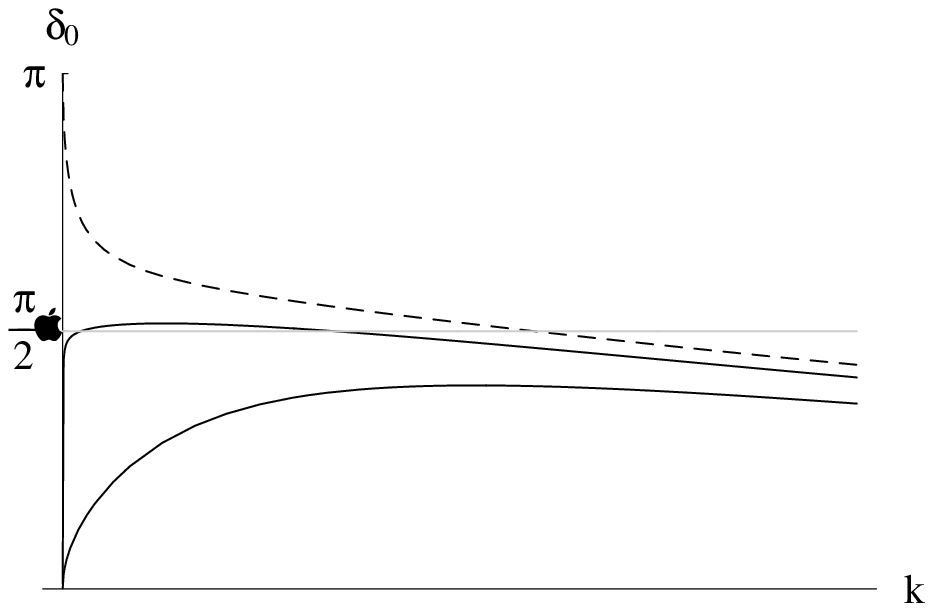} \medskip
\caption{The s-wave phase shifts for a Dirac particle in a spherical well. The critical potential $V_c=4.20m$. The solid lines represent potentials of strength $V=4.195m$ (middle line) and $V=4m$ (lowest line). The dashed line is a potential well of strength $V=4.3m$ which contains a bound state. ($a=1/m$). }
\end{center}
\end{figure}
It can be seen that the s-wave phase shift now just increases through $\frac{\pi}{2}$ for values of the potential that are weaker than a critical value $V_c$ and thus we have a very weak resonance by our definition above. As will be shown below, the width of this resonance is large compared to that for the p-wave. This is a new result - we would not expect to obtain resonances for s-waves in a spherical potential well. 

The p-wave phase shifts are very similar to those obtained for p-waves in the Schr\"odinger case and are illustrated below.
\begin{figure}[tbph]
\par
\begin{center}
\leavevmode
\includegraphics[width=0.4\linewidth]{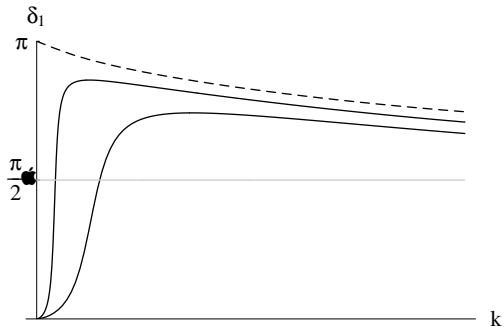} \medskip
\caption{The p-wave phase shifts for a Dirac particle in a spherical well. The critical potential $V_c=2.30m$. The solid lines represent potentials of strength $V=2m$ (lowest line) and $V=2.2m$ (middle line). The dashed line is a potential well of strength $V=2.4m$ which contains a bound state. ($a=1/m$). }
\end{center}
\end{figure}
For energies close to the resonant energy position, $E_R$, the resonance has the Breit-Wigner form:
\begin{equation}
\tan \delta_{l_\chi} \simeq \frac{\Gamma_{l_\chi}/2}{E_R-E}
\label{res63}
\end{equation}
where $\Gamma_{l_\chi}$ is the width. This width is related to the time delay \cite{Landau} by
\begin{equation}
\Gamma_{l_\chi} \simeq \frac{1}{\tau_{l_\chi}}
\label{res64}
\end{equation}
By taking the ratio of the widths of the s-wave resonance for $V=4.195m$ and the p-wave wave resonance for $V=2.25m$, the s-wave resonance is seen to be approximately $230$ times wider.

For a very clear illustration of the existence of s-wave resonances for the relativistic spherical well, we use the values for the well width $a$ and depth $V$ of Pieper and Greiner \cite{PG} in their discussion of superheavy nuclei. In that paper they establish the supercritical values for s states but do not consider scattering phase shifts. Here we choose $a$ to $8$ fm and then the first three critical values at $E=m$ and $E=-m$ for s and p states are easily calculated using Eqs. (\ref{res50}, {\ref{res51}, \ref{res55} and \ref{res57}) and are tabulated below:

\begin{center}
\begin{tabular}{|c|c|c|c|}\hline
\quad s (E=m) &\quad s (E=-m) \quad&\quad p (E=m) \quad&\quad p (E=-m) \quad \\ \hline
\quad 75.947 \quad&\quad  77.997 \quad&\qquad  76.975 \quad&\quad  79.012 \quad\\
\quad 153.434 \quad&\quad  155.480 \quad&\qquad  154.458 \quad&\quad  156.498 \quad\\
\quad 230.919 \quad&\quad  232.964 \quad&\qquad  231.942 \quad&\quad  233.983 \quad\\
\hline
\end{tabular}
\par~\par Table~1:~The critical well depths in MeV for s- and p- states at $E= \pm m$.
\end{center}

The critical well for the $1s_{\half}$ state is now weakened slightly by 1\% to make the depth $V=75.187$ MeV. The phase shift is seen to be: 

\begin{figure}[tbph]
\par
\begin{center}
\leavevmode
\includegraphics[width=0.8\linewidth]{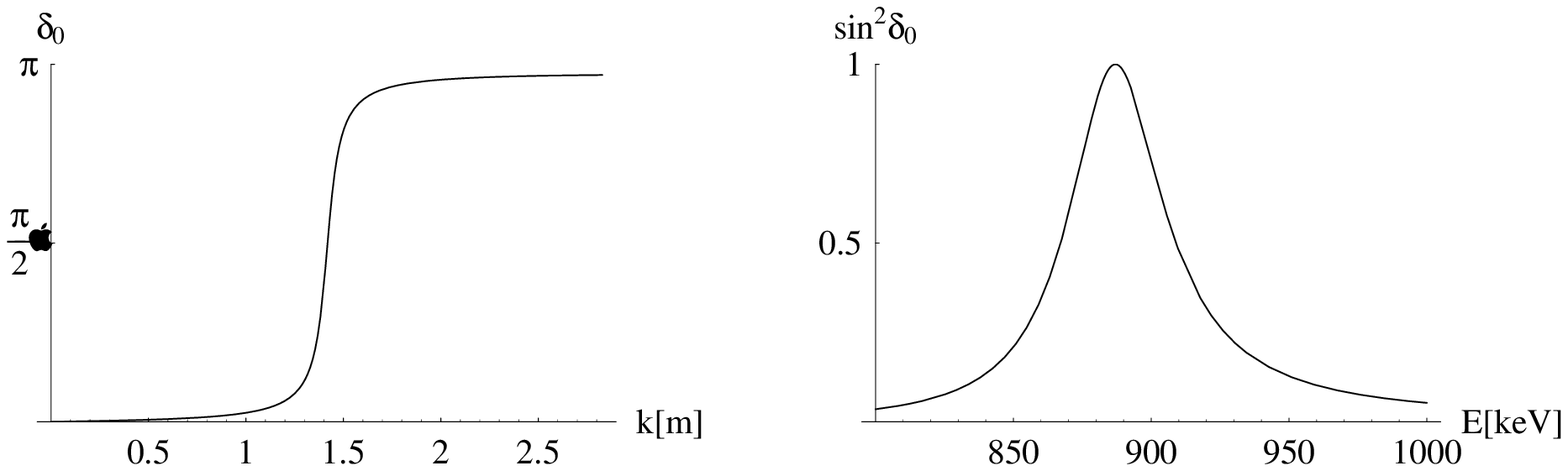} \medskip
\caption{The $1s_{\half}$ phase shift and energy dependence of $\sin^2 \delta_0$ for the spherical well ($a=8$ fm and $V=75.187$ MeV).}
\end{center}
\end{figure}
The phase shift is clearly seen to increase through $\frac{\pi}{2}$ corresponding to a resonance. The energy dependence of $\sin^2 \delta_0$ is plotted beside the phase shift and the width is found to be $\Gamma_0=41.83$ keV.

Similarly the critical well for the $2p_{\half}$ state is weakened slightly by 1\% to make the depth $V=76.205$ MeV. The phase shift and energy dependence of $\sin^2\delta_1$ is seen to be: 

\begin{figure}[tbph]
\par
\begin{center}
\leavevmode
\includegraphics[width=0.8\linewidth]{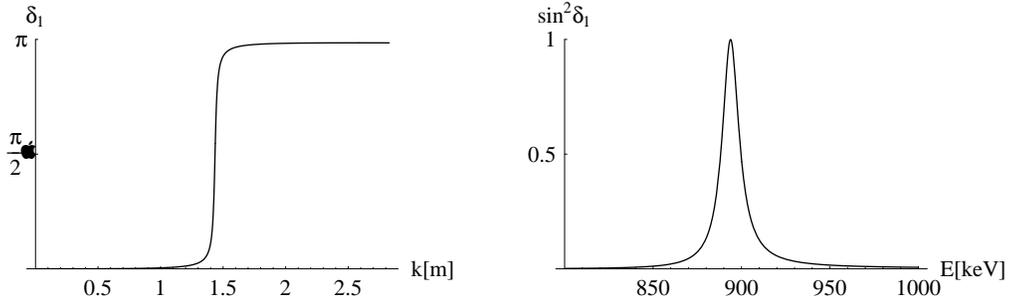} \medskip
\caption{The $2p_{\half}$ phase shift and energy dependence of $\sin^2 \delta_1$ for the spherical well ($a=8$ fm and $V=76.205$ MeV).}
\end{center}
\end{figure}

The width of the $2p_{\half}$ resonance is seen to be $\Gamma_1=11.5$ keV. The $2p_{\half}$ phase shift is seen to increase more rapidly through $\frac{\pi}{2}$ than the $1s_{\half}$ phase shift and consequently the $1s_{\half}$ resonance is wider than the $2p_{\half}$ resonance by a factor of approximately $3.6$.

In order to understand why s-wave resonances occur for relativistic particles we de-couple the large, $f$, and small, $g$, components of Eqs. (\ref{res13}) to form two second order equations. In the case of a general short-range potential the equations for solutions outside the potential are: 
\begin{eqnarray}
g^{\prime \prime}-\left[E^2-m^2-\frac{\chi(\chi+1)}{r^2} \right]g(r)&=&0 \nonumber\\
f^{\prime \prime}-\left[E^2-m^2-\frac{\chi(\chi-1)}{r^2} \right]f(r)&=&0 
\label{res65}
\end{eqnarray}
What happens is that the angular momentum barrier for the large component vanishes for $s_{\half}$-waves when $\chi=-1$ but the small component still has an angular momentum term. In contrast for $p_{\half}$-waves ($\chi=1$), the barrier remains for the large component but disappears for the small component so the centrifugal barrier never vanishes for both $f$ and $g$. With this angular momentum term in one or more of the components we are considering cases analogous to the $l>0$ type scattering in the Schr\"odinger equation and hence this gives rise to real resonances.


\section*{Phase Shift Behaviour : The Spherical Barrier}

In the case of low momentum scattering $E \to m$ on the spherical barrier, the analysis of the previous section can be repeated with $V \to -V$. The barrier momentum is now $p \to q =\sqrt{(E-V)^2-m^2}$. In the limit $E \to m$ this becomes $q \to p_{-}$. The corresponding form of Eq. (\ref{res59}) for scattering from barriers is
\begin{equation}
\frac{j_{-\chi-1}(p_{-}a)}{j_{-\chi}(p_{-}a)}=
\frac{2ma}{1+2\chi} \sqrt{\frac{V}{V-2m}}
\label{res66}
\end{equation}
which is the supercritical result for $\chi>0$ given by Eq. (\ref{res50}). Similarly Eq. (\ref{res60}) becomes
\begin{equation}
j_{\chi-1}(p_{-}a)=0
\label{res67}
\end{equation}
which is the supercritical result for particles with $\chi<0$ given by Eq. (\ref{res44}).

In the previous section we showed that the conditions for s-wave resonance scattering at low momenta off a spherical well agreed with the critical conditions at $E=m$ calculated from the bound s state spectrum of that well. The same is true for p waves. However when s-waves are incident on a barrier this occurs for critical potentials which correspond to the supercritical values for p-waves in the spherical well (i.e. when the p-wave is at $E=-m$). Similarly scattering p-waves are in resonance when the s-wave is supercritical. This phenomenon is known as crossing and was first discussed by two of us \cite{DH} in a discussion of positron scattering from heavy nuclei. Under the transformation $E \to -E$, $V \to -V$, $\chi \to -\chi$, $f \to g$, the coupled equations (\ref{res13}) are seen to be invariant. Consequently a supercritical $2p_{\half}$ state for a well corresponds to a resonant $1s_{\half}$ state at $E=m$ on a barrier. 

In order to calculate the $1s_{\half}$ phase shift for scattering from a barrier we must use this result to calculate the supercritical condition for $2p_{\half}$ particles. By strengthening the supercritical well slightly by $1$\%, which, through crossing, increases the height of the barrier, the $1s_{\half}$ phase shift and partial cross-section are seen to be: 
\begin{figure}[tbph]
\par
\begin{center}
\leavevmode
\includegraphics[width=0.8\linewidth]{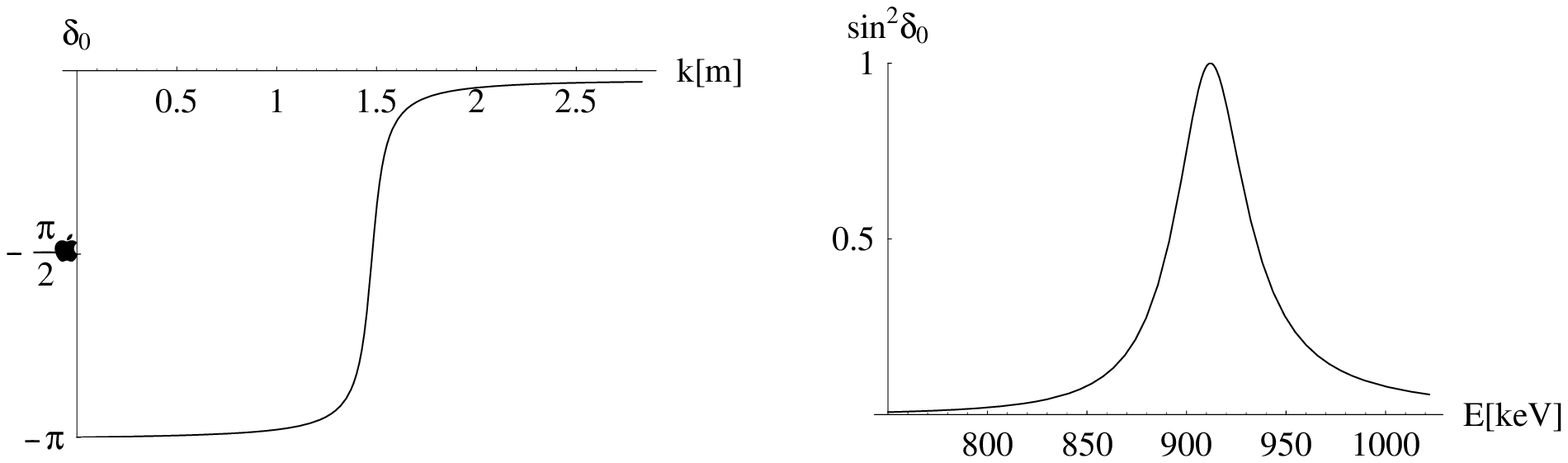} \medskip
\caption{The $1s_{\half}$ phase shift and energy dependence of $\sin^2 \delta_0$ for the spherical barrier ($a=8$ fm and $V=79.802$ MeV) .}
\end{center}
\end{figure}
We see a very clear s-wave resonance with width $\Gamma_0=43.1$ keV. Similarly if the supercritical well for the $1s_{\half}$ state is strengthened by 1\% the phase shift for scattering from the barrier $V=78.777$ MeV is:

\begin{figure}[tbph]
\par
\begin{center}
\leavevmode
\includegraphics[width=0.8\linewidth]{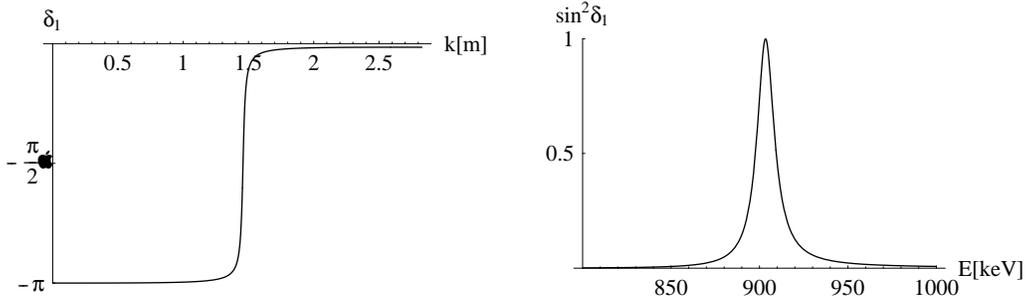} \medskip
\caption{The $2p_{\half}$ phase shift and energy dependence of $\sin^2 \delta_1$ for the spherical barrier ($a=8$ fm and $V=78.777$ MeV).}
\end{center}
\end{figure}
The width of the $1s_{\half}$ resonance is seen to be $\Gamma_1=11.8$ keV and the ratio of widths of the $1s_{\half}$-wave to the $2p_{\half}$-wave is found to be $3.7$.


\section*{General Potentials}

The reasoning we have used to understand the analytical results obtained for the spherical well
applies equally to monotone short-range potentials with more general shapes that vanish at infinity.  We shall suppose that the potential $V(r)$ is spherically symmetric and is given by
\begin{equation}
V(r)= \pm\ v w(r/a)
\label{res68}
\end{equation}
\noindent where $v > 0$ is a depth parameter, $a$ is a range parameter, $x = r/a > 0$ is a dimensionless distance, and $w(x)$ is a dimensionless shape function. Thus $\ w(x) = e^{-x^2}$ represents a Gaussian potential and $w(x) = 1,\ x < 1;\ w(x) = 0,\ x \geq 1$ is the spherical well which we have already studied analytically.  Clearly we cannot always expect to find exact analytical solutions.  Numerical methods are flexible and powerful but involve arbitrary features, such as iteration method, step size, and boundary at `infinity', which introduce corresponding uncertainties. Our approach is therefore to build a programme, a kind of `virtual laboratory', to test it on the spherical well, and then to apply it to the Gaussian potential.  For $V(r) < 0,$ this programme would have to be able to find discrete eigenvalues; for potentials of either sign, it would need to be able to compute the wave-function amplitudes at the origin resulting from spherical waves of unit amplitude which are directed towards the origin from far away.  Since the scattering process is, in principle, reversible, we shall actually start waves with amplitude $C$ at the origin, and then adjust $C$ so that the waves at distant points have unit amplitude.  In addition, we shall need to define and compute a `numerical phase (shift)'. This question has been discussed in detail in reference \cite{IRS}. 

We shall now explain more fully what we mean in this discussion by the terms `amplitude' and `numerical phase'. For very small values of $r,$ the potentials we study are essentially indistinguishable from spherical wells or barriers. Hence the boundary conditions at the origin are the same as those for the exact spherical well solutions (Eq. \ref{res14}). We shall find it convenient to separate the sign and the magnitude of $\chi$ by use also of the symbols $k = |\chi| = j+\half,$ and $\tau = \chi/|\chi| = \chi/k.$ In the small-$r$ region we may write
\noindent for $\chi > 0$
\begin{eqnarray}
f &=& Cr^{k+1}\nonumber \\
g &=& C' r^{k} = \sigma C r^{k}\nonumber \\
\frac{rg}{f} &=& \frac{2k+1}{E+m-V} = \frac{C'}{C} = \sigma\nonumber \\
\sqrt{f^2+g^2} &=& |C| r^k \sqrt{r^2 + \sigma^2}\approx \sigma |C| r^k,\quad r \ll 1
\label{res69} 
\end{eqnarray}

\noindent and for $\chi < 0$

\begin{eqnarray}
f &=& C r^{k}\nonumber \\
g &=& C'r^{k+1} = \sigma C r^{k+1}\nonumber \\
\frac{g}{rf} &=& \frac{V-E+m}{2k+1} = \frac{C'}{C} = \sigma\nonumber \\
\sqrt{f^2+g^2} &=& |C| r^k \sqrt{1 + r^2 \sigma^2}\approx |C| r^k,\quad r \ll 1.
\label{res70}
\end{eqnarray}

\noindent Ignoring the $fg$-ratio factor $\sigma,$ we shall say of these cases that the amplitude at the origin is $C.$  We therefore seek the value of $C$ which yields $\sqrt{f^2+g^2} = 1$ as $r\rightarrow \infty.$  We recall that the `radial functions' $f(r) = r\psi(r)$ and $g(r) = r\phi(r)$ each have in their definitions a factor $r;$ for bound states we have (omitting the angular factor $4\pi)$ the corresponding normalisations,
\begin{equation}
\int_0^{\infty}\sqrt{f^2(r) + g^2(r)}dr 
= \int_0^{\infty}\sqrt{\psi^2(r) + \phi^2(r)}r^2 dr = 1.
\label{res71}
\end{equation}
\noindent If we consider the case $j = 1/2\Rightarrow k = 1,$ then we have for $\chi > 0$ $\sqrt{\psi^2(0) +\phi^2(0)} = \sigma |C|,$ and for $\chi < 0$ $\sqrt{\psi^2(0) +\phi^2(0)} = |C|.$ This example explains our use of the phrase 'amplitude at the origin' to describe $C.$

At large distances the asymptotic scattering solutions are essentially sinusoidal, like those of the spherical barrier (Eqs. \ref{res26},\ref{res27}). In this region, we choose a certain node $\nu$ in $g$, thus $g(r_{\nu})=0,$ and the amplitude $\sqrt{f^2+g^2}$ at this point is given by $|f(r_{\nu})|.$ Since for the Gaussian potential, the potential value $V(r)$ is never zero, we must choose a sufficiently far node that the `sinusoidal' asymptotic region has essentially been reached.  For the spherical well and Gaussian potential we have found $\nu = 20$ to be satisfactory. Our task then is to fix $\nu$ and then determine $C$ so that $|f(r_{\nu})| = 1.$  Equivalently, we can set $C = 1,$ and then use the resulting $|f(r_{\nu})|^{-1}$ as the value of $C$ sought. It may be helpful at this point to see the wave functions in a particular case: they are shown in Figure~\ref{resfig} for the scattering of $s_{\half}$-state waves of momentum $p = 0.1$ by a Gaussian barrier. 
\begin{figure}[htp]
     \centering
     \subfigure[The large ($f$) and small ($g$) components for $r\leq7$.]{
          \label{resfig:figa}
          \includegraphics[width=.45\linewidth]{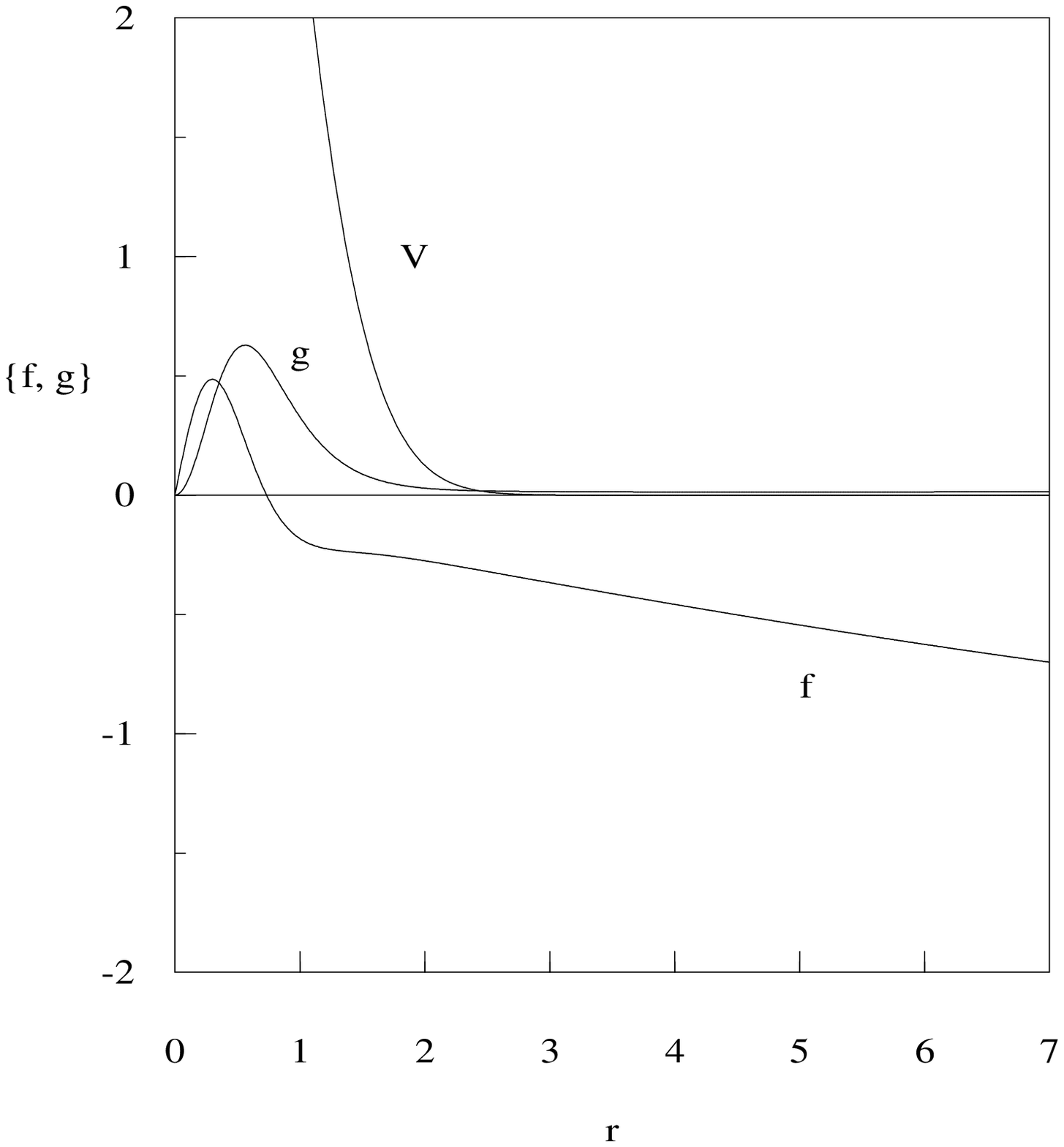}}
     \subfigure[The {\lq}large component' $f$ has a maximum of $1$ near the 20th node $r_{20}=635.2$ of the {\lq}small component' $g$.]{
          \label{resfig:figb}   
          \includegraphics[width=.45\linewidth]{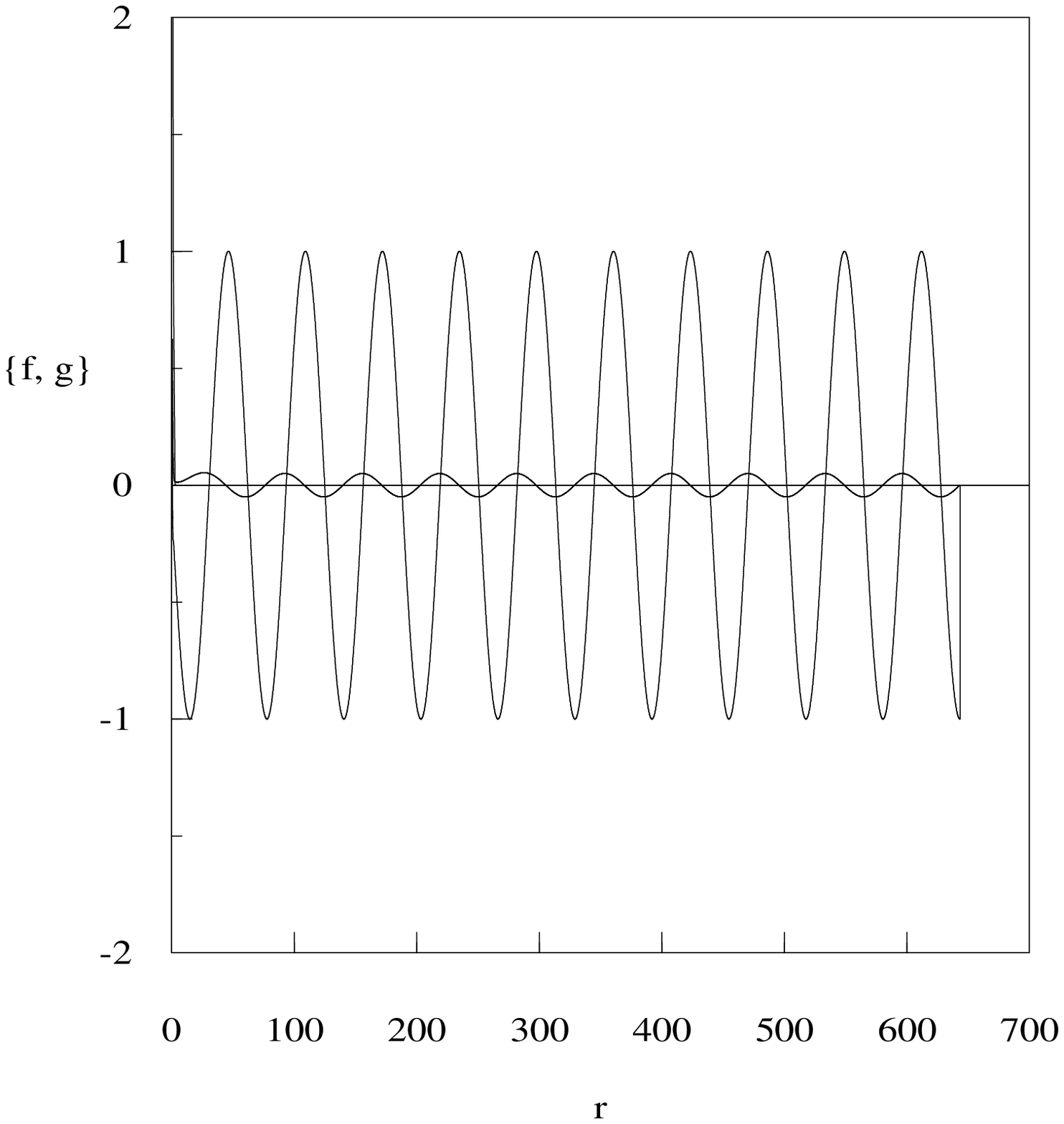}}\\

     \caption{Resonant s-wave wave function components for a particle with incident momentum $p=0.1$ scattering from a Gaussian barrier $V(r)=v \, \exp(-r^2)$, $v=6.8$.}
     \label{resfig}
\end{figure}
The node $r_{20}$ of $g$ has been detected at $r_{20} = 635.2$, and the corresponding maximum of $f$ has the value $1$ with the choice $C = 12.8.$  The final determination of the maximum is achieved by fitting a sine function to $f(r)$ locally, as will now be explained.

The position of the maximum allows us at the same time to determine what we shall call the `numerical phase' $\delta.$  We use a modular function to compare the position of the peak with respect to a grid.  The details of our function have some arbitrary elements which have been designed to yield a suitably scaled indicator of the phase.  We start with an explicit model of the wave function for the large-$r$ asymptotic region.  The step size of our numerical integration procedure is represented by $h.$ The $\nu^{\rm th}$ node $r = r_{\nu}$ of $g$ indicates the approximate position of the maximum in $f.$  We write  
\begin{equation}
f(r) = \alpha = D\sin(pr+\delta'), \quad \beta = f(r+h),\quad \theta = pr+ \delta'.  
\label{res72}
\end{equation}  
\noindent Provided $\alpha \ne \beta$ and $\sin(\theta)\ne 0$ (special equations are required for the rare cases of equality), we find
\begin{equation}
\theta = \tan^{-1}\left(\frac{\sin(p h)}{\cos(ph) -\beta/\alpha}\right),\quad D = \frac{\alpha}{\sin(\theta)}.
\label{res73}
\end{equation}
\noindent We now define $\delta$ by the equations
\begin{equation}
\delta_1 = \theta - pr + (1-\tau)\frac{\pi}{4},\quad \delta = \delta_1 - \pi\lfloor\delta{_1}/\pi\rfloor + \frac{\pi}{2},
\label{res74}
\end{equation}

\noindent where we have used $\lfloor x\rfloor$ to represent the integer part of $x$ (often written as ${\rm int}(x)$ in computer languages). This definition of $\delta$ yields a number in the range $\delta \in [-\frac{\pi}{2}, \frac{\pi}{2}].$  We have added a plot of $\delta(v)$ to the top of our resonance graphs of $C(v),$ presented below.  Eq. (\ref{res73}) also indicates explicitly how $C(v)$ is found: the value of $v$ is chosen; the value of $C$ in Eq. (\ref{res70}) is set to 1; the value of $D$ is determined by Eq. (\ref{res73}); then we have $C(v) = D^{-1}.$ In order to plot Figure~2b, the value of $C$ in Eq. (\ref{res70}) is now set to $C(v)$ just found, and this leads to $D = 1,$ as the graph shows.  All this computational activity is packaged in the form of a C++ class called `dirac'. An instance `{\bf dc}' of dirac is created and messages are sent to {\bf dc} to set the potential and the coupling $v$ and the initial conditions etc; {\bf dc} is then requested to compute eigenvalues and to perform scattering `experiments', and to report the results.

This scattering quantity can be written in a closed form for the spherical potential. We start by looking at the asymptotic form of the wave function for very large values of $r$ given by Eqs. (\ref{res26} to \ref{res30}), it is possible to find a value of $r=R$ where the wave function has a peak (or node) in the top component and a corresponding node (or peak peak) in the bottom component. One such set of values is given by
\begin{equation}
R=\frac{1}{k}\left(N \pi+l_{\chi}\frac{\pi}{2}-\delta_{l_{\chi}}\right)
\label{res75}
\end{equation}
where $N$ is an arbitrarily large integer. From this choice the $f$ component will exhibit a node whilst the $g$ component will be at a peak or trough of the sinusoidal wave. The modulus of the wave function at $r=R$ from Eqs. (\ref{res26} and \ref{res27}) is seen to be
\begin{equation}
|\Psi(R)|=\frac{A}{E+m}=\frac{\sqrt{b_1^2+b_2^2}}{E+m}
\label{res76}
\end{equation}
$\rho$ is a measure of the amplitude of the wave function at the origin when the modulus of the incident wave at great distances is unity. It is therefore important to examine the form of the wave function very close to the origin. The components of the wave function inside the potential have the form given by Eqs. (\ref{res14}). Using the following identity for the limiting value of the spherical Bessel function for small arguments
\begin{equation}
\lim_{x \to 0} \left[x^{-n}j_n(x) \right] =\frac{1}{(2n+1)!!}
\label{res77}
\end{equation}
\begin{equation}
\Psi(r) =\left(
\begin{array}{c}
f(r)\\
g(r)
\end{array}
\right)
\to a_1 \left( \begin{array}{c}
\frac{r(pr)^{l_\chi}}{(2l_\chi+1)!!} \\
\frac{\chi}{|\chi|}\frac{1}{E-V+m} \frac{(pr)^{l_{-\chi}+1}}{(2l_{-\chi}+1)!!}
\end{array} \right)
\label{res78}
\end{equation}
The modulus of the wave function at the origin is therefore
\begin{eqnarray}
|\Psi(0)| =\sqrt{f^2(0)+g^2(0)}\simeq \left\{
\begin{array}{cl}
a_1 \displaystyle{\frac{p^{\chi}}{E-V+m}}\displaystyle{\frac{1}{(2\chi-1)!!}}r^{\chi} &\qquad \chi>0\\[0.3cm]
a_1 \displaystyle{\frac{p^{-\chi-1}}{(-2\chi-1)!!}}r^{-\chi}&\qquad \chi<0
\end{array}\right.
\label{res79}
\end{eqnarray}
In particular for s-waves
\begin{equation}
|\Psi(0)|  \simeq a_1 r
\label{res80}
\end{equation}
whilst for p-waves
\begin{equation}
|\Psi(0)|  \simeq  a_1 \frac{pr}{E-V+m}
\label{res81}
\end{equation}
in agreement with Eqs. (\ref{res69} and \ref{res70}). Following the argument given above we can set $a_1=1$ and the quantity $C(\pm)$ - where the sign indicates a barrier (+) or well (-) - is found by
\begin{equation}
C(\pm)=\left|\frac{1}{\Psi(R)}\right|
\label{res82}
\end{equation}
This can be established upon substitution of Eqs. (\ref{res76}, \ref{res23} and \ref{res24}) into the above. $C(-)$ can be established from the above by the transformation $V \to -V$.

\par For $m = 1$ we decided first to study three resonance curves, that is to say, graphs of $C(v)$ for a range of $v>0$ large enough to reach just beyond the third resonance (this is illustrated for the Gaussian potential).  We kept $j = \half$ constant and studied the four cases: $\tau = -1$ ($s_{\half}$), $\tau = 1$ ($p_{\half}$), repulsive barrier $(+),$ and attractive well $(-).$  For both of the potentials we found that the resonances associated with the attractive wells corresponded exactly to the critical eigenvalues at $E = m$ in the same sector, $s$ or $p;$  whereas, for scattering from barriers, the resonances corresponded, in agreement with our crossing theorem, to the eigenvalues $E = -m$ in the `other' sector, $s$ to $p,$ and $p$ to $s.$ For the spherical potential, the exact critical couplings are given by the following formulas in the \hi{p}-{sector}:
\begin{eqnarray}
p_{\half}(-)\quad&\Rightarrow&\quad v = \left(1+ (n\pi)^2\right)^{\half}-1,\quad n = 1,2,3,\dots
\label{res83} \\
p_{\half}(+)\quad&\Rightarrow&\quad v = \left(1+ (n\pi)^2\right)^{\half}+1,\quad n = 1,2,3,\dots
\label{res84}
\end{eqnarray}
\noindent which can be established from Eqs. (\ref{res47} and \ref{res51}) with $a=1$. The corresponding values in the \hi{s}{sector} are found as solutions (Eqs. \ref{res46} and \ref{res53}) to a quasi-bound state problem in which the spinor is not normalised since $\phi(r)$ approaches a constant non-zero value as $r\rightarrow \infty.$ The numerical values relevant to the resonance curves are shown in Table~1. 

\begin{center}
\begin{tabular}{|c|c|c|c|}\hline
$s_{\half} (+)$ & $s_{\half} (-)$   & $p_{\half} (+)$     & $p_{\half} (-)$\\ \hline
 5.27& 1.11 & 4.30 & 2.30    \\
8.40 & 4.20 & 7.36 & 5.36  \\
11.54 & 7.33 & 10.48 & 8.48 \\ \hline
\end{tabular}
\par~\par Table~2:~Critical couplings for scattering from spherical barriers $(+)$ and wells $(-)$.
\end{center}
\noindent  As we have explained, these critical couplings correspond to $(V < 0)$ eigenvalues $E = \pm m$ by the rules:
\begin{eqnarray}
s_{\half}(+)&\leftrightarrow &(E = -m)p_{\half}, 
\label{res85} \\
p_{\half}(+)&\leftrightarrow &(E = -m)s_{\half},
\label{res86} \\
s_{\half}(-)&\leftrightarrow &(E = m)s_{\half}, 
\label{res87} \\
p_{\half}(-)&\leftrightarrow &(E = m)p_{\half}.
\label{res88}
\end{eqnarray}

The corresponding $C(p)$ resonance curves are shown in Figure~\ref{resfig2}. 

\begin{figure}[htp]
     \centering
     \subfigure[$C(p)$ for s-wave scattering from a spherical barrier of height $U=5.29$.]{
          \label{resfig2:figa}
          \includegraphics[width=.35\linewidth]{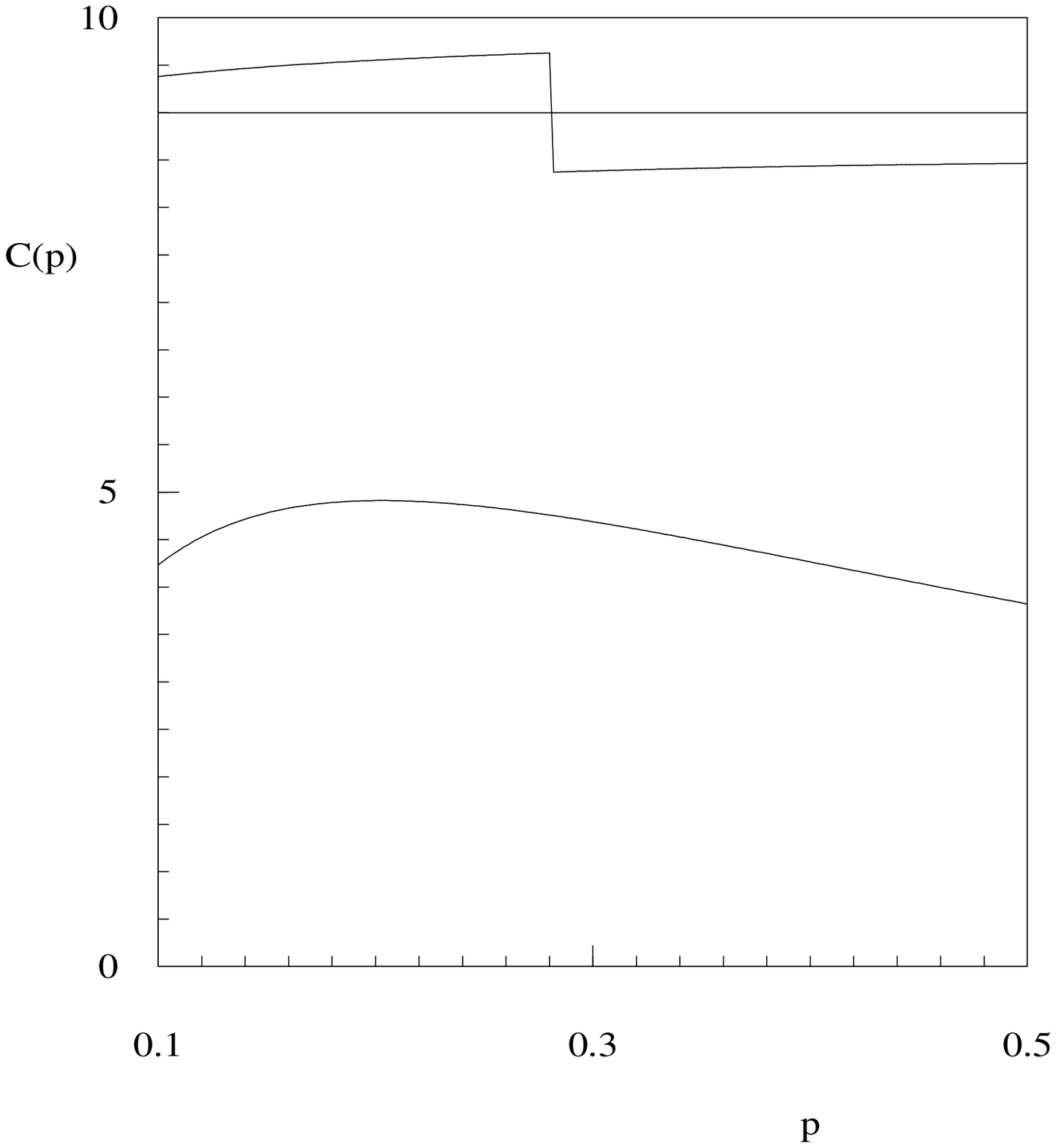}}
     \hspace{.1in}
     \subfigure[$C(p)$ for p-wave scattering from a spherical barrier of height $U=4.4$.]{
          \label{resfig2:figb}   
          \includegraphics[width=.35\linewidth]{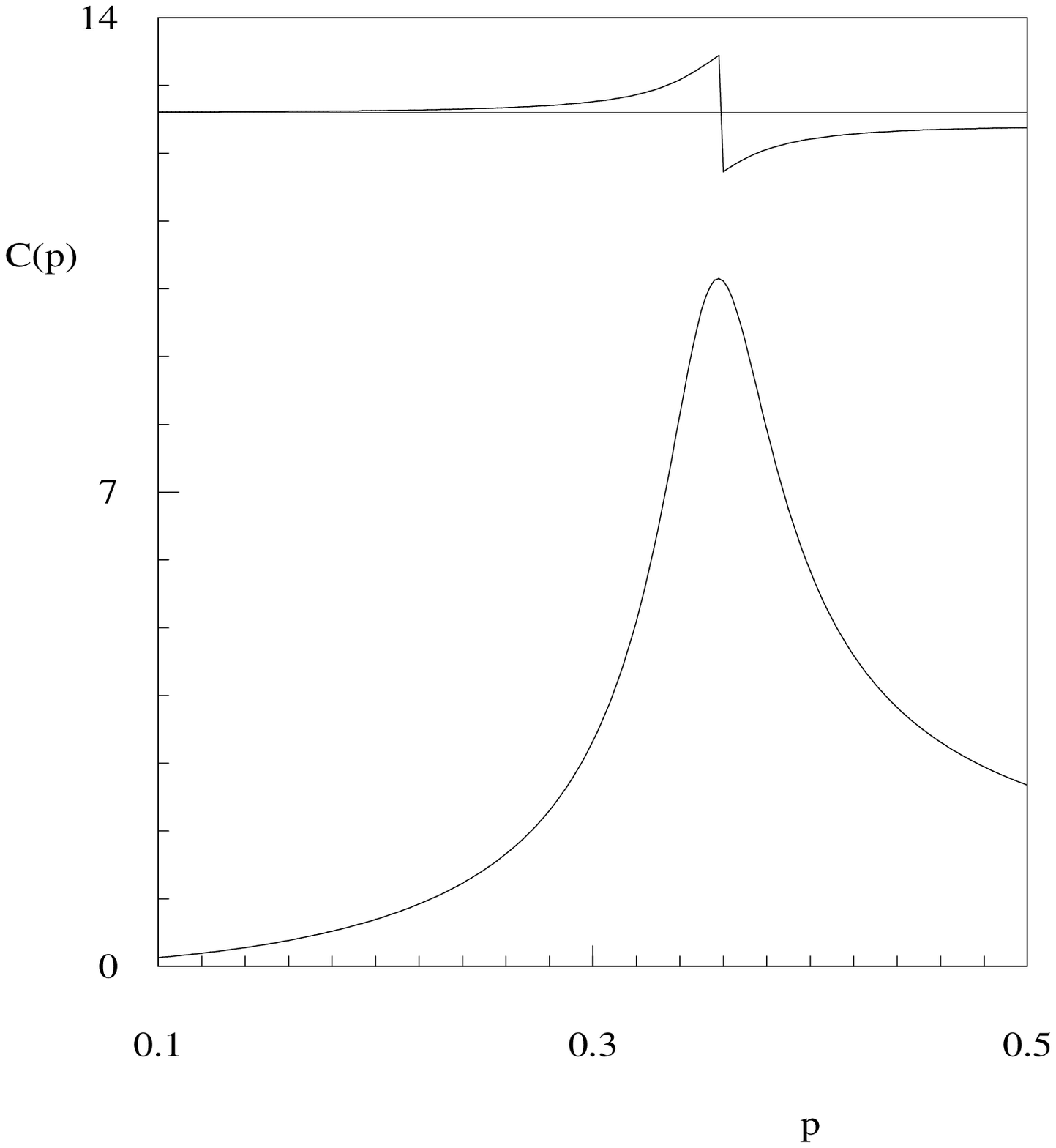}}\\
     \vspace{.1in}
     \subfigure[$C(p)$ for s-wave scattering from a spherical well of depth $U=-4.195$.]{
           \label{resfig2:figc}
           \includegraphics[width=.35\linewidth]{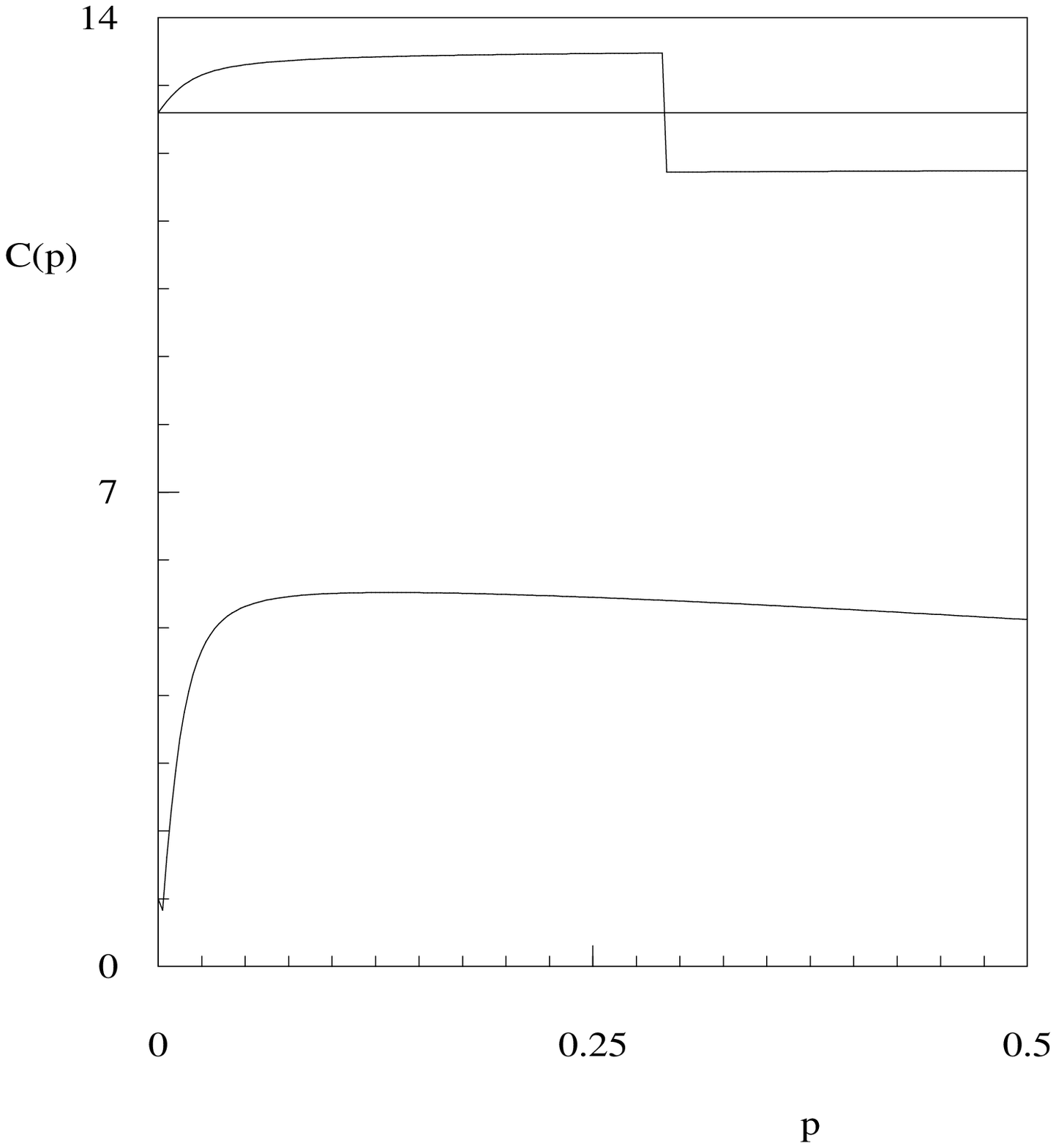}}
     \subfigure[$C(p)$ for p-wave scattering from a spherical well of depth $U=-2.2$.]{
           \label{resfig2:figd}
          \includegraphics[width=.35\linewidth]{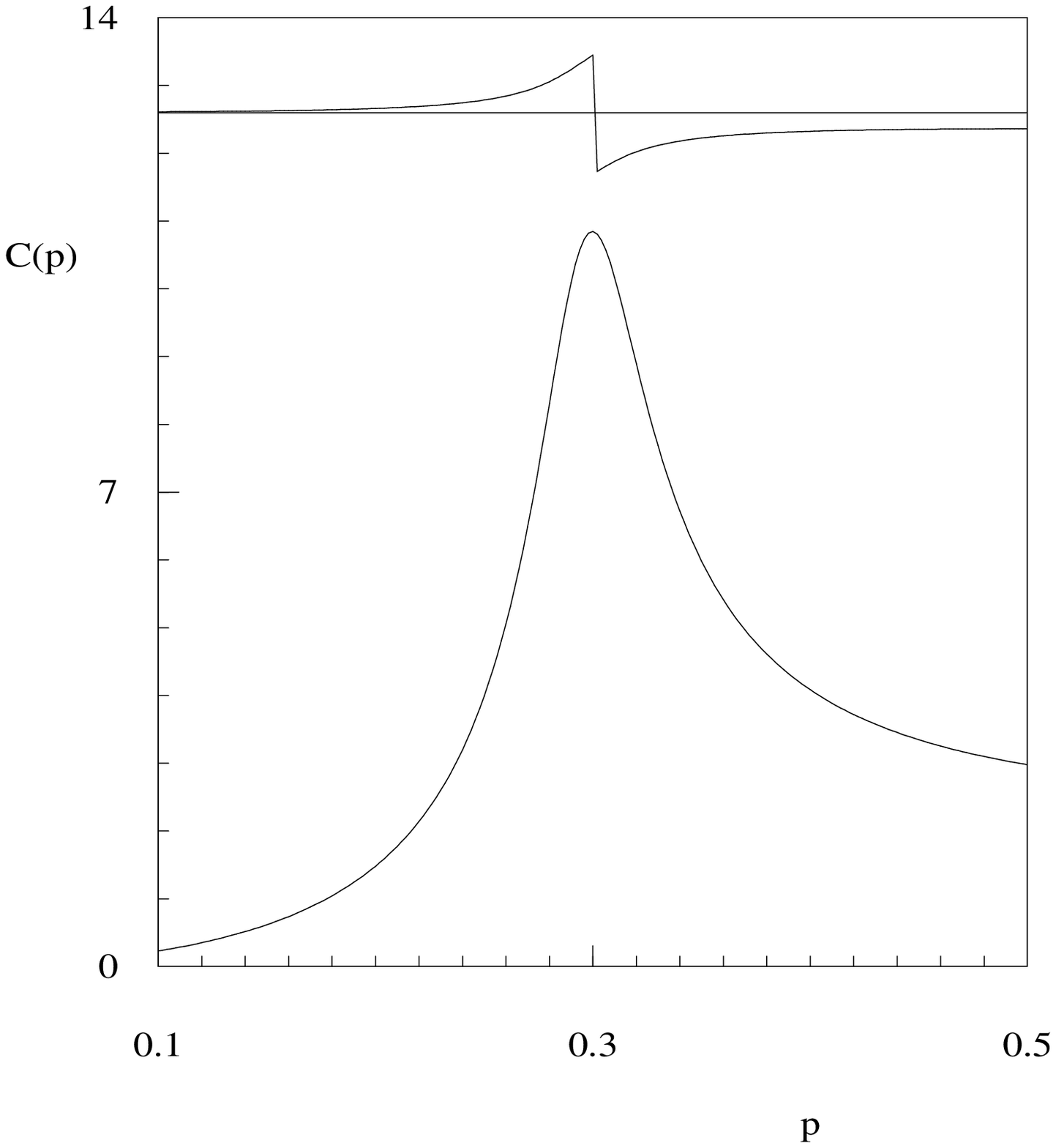}}
     \caption{These plots show the scattering measurement $C(p)$ plotted against the incident 
momentum $p$ for $s$ and $p$ waves incident on supercritical spherical barriers and subcritical wells.}
     \label{resfig2}
\end{figure}

By repeating the analysis for the Gaussian potential, we obtain the resonance peaks and corresponding eigenvalues $E = \pm\ m$ at the critical couplings shown in Table~2. 

\begin{center}
\begin{tabular}{|c|c|c|c|}\hline
$s_{\half} (+)$ & $s_{\half} (-)$ & $p_{\half} (+)$ & $p_{\half} (-)$\\ \hline
6.75 & 1.26 & 5.62 & 2.96\\ 
10.42 & 4.37 & 9.23 & 6.11\\ 
14.04 & 7.70 & 12.83 & 9.44\\ \hline
\end{tabular}
\par~\par Table~3: Critical couplings for Gaussian barriers $(+)$ and wells $(-).$
\end{center}

\noindent The resonance curves $C(v)$ for an incident momentum of $p = 0.1$ for the Gaussian potential are shown in Figure~\ref{resfig3}. We have also computed the critical couplings corresponding to $s$ and $p$ states with $E = -m = -1$ and $E = m = 1.$ Allowing for $s\leftrightarrow p$ crossing in the $E = -m$ cases, 
these values agree with the 
scattering results reported in Table~1. The heights of the peaks in these resonance curves are somewhat arbitrary:  they depend on the choice of $p,$ and, more randomly, on the number of $v$ values used in the plot. Ideally we should have $p\rightarrow 0:$ the barrier peaks are sharper for smaller $p,$ and their heights become unbounded; these effects can be `seen' numerically by  fine explorations of $C(v)$ for $v$ near critical values.  However, as $p$ is reduced, the scale of the relevant part of the wave function expands, and the numerical integration must be
carried out to greater and greater distances.  The choice $p = 0.1$ and its implications are therefore arbitrary.  Moreover, the plots we exhibit were made with $n_v = 400$ points in the abscissae: the closeness to which the resulting values of $v$ lie to critical values is therefore a matter of chance.   The resulting numerical maxima have these arbitrary aspects, which, of course, emphasise, by comparison, the great importance of the study of exact analytical solutions when such are available.

\begin{figure}[htp]
     \centering
     \subfigure[s-wave resonances and phase shift for the Gaussian barrier.]{
          \label{resfig3:figa}
          \includegraphics[width=.35\linewidth]{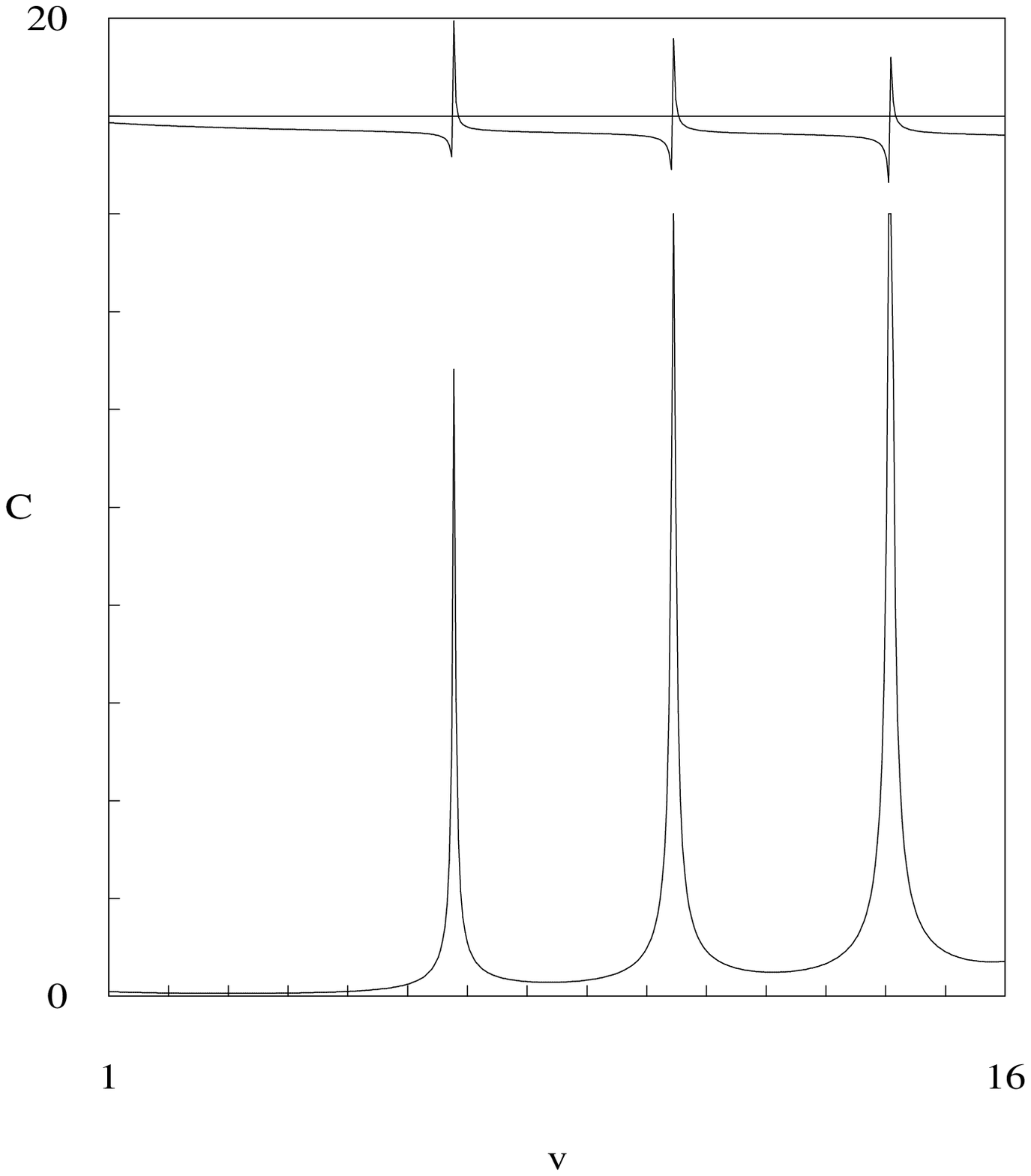}}
     \hspace{.1in}
     \subfigure[p-wave resonances and phase shift for the Gaussian barrier.]{
          \label{resfig3:figb}   
          \includegraphics[width=.35\linewidth]{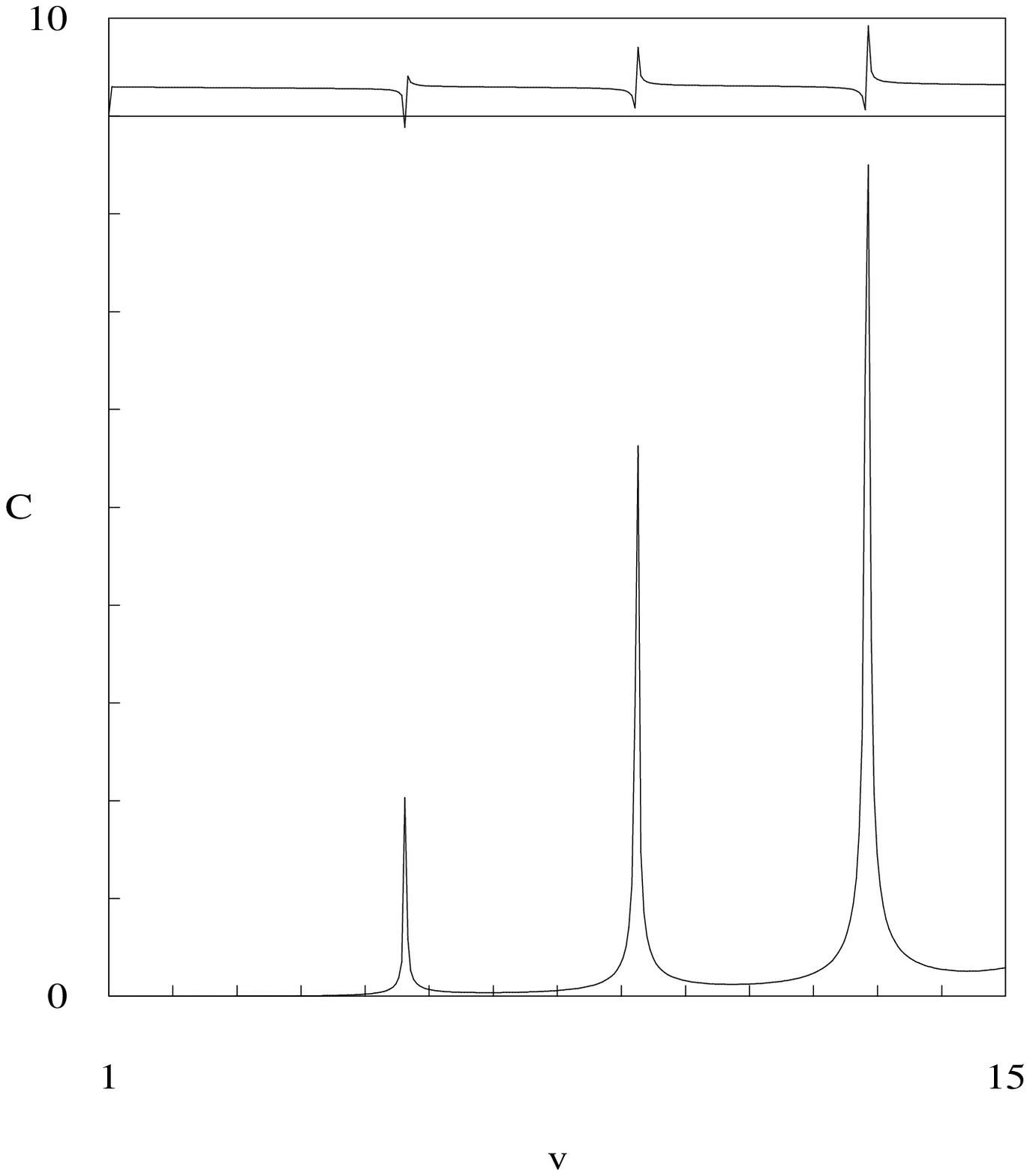}}\\
     \vspace{.1in}
     \subfigure[s-wave resonances and phase shift for the Gaussian well.]{
           \label{resfig3:figc}
           \includegraphics[width=.35\linewidth]{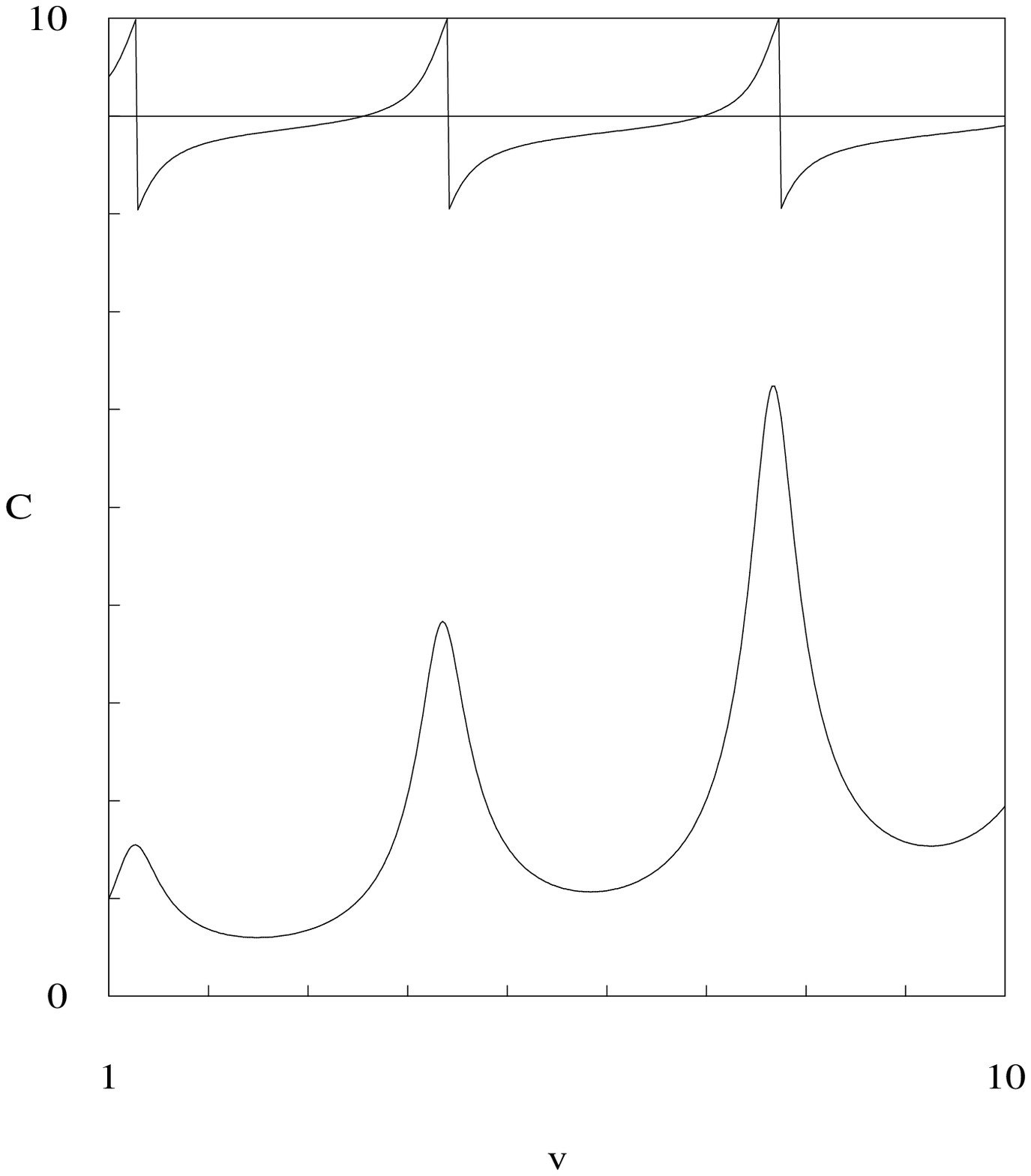}}
     \subfigure[p-wave resonances and phase shift for the Gaussian well.]{
           \label{resfig3:figd}
          \includegraphics[width=.35\linewidth]{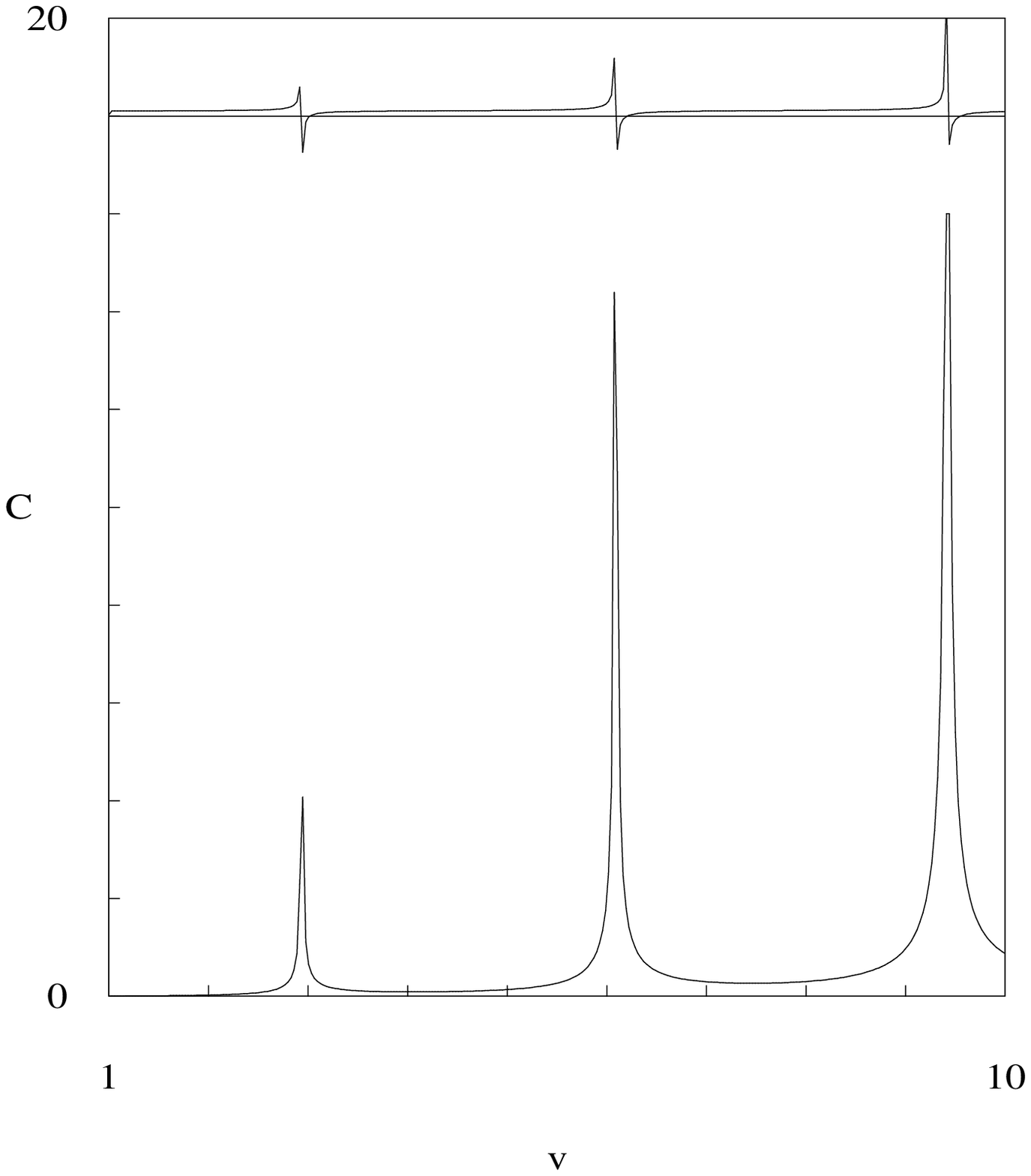}}
     \caption{These plots show the scattering measurement $C(p)$ plotted against the incident 
momentum $p$ for $s$ and $p$ waves incident on supercritical Gaussian barriers and subcritical wells.}
     \label{resfig3}
\end{figure}


We have found similar results for other short-range potential shapes, such as the exponential potential $w(x) = e^{-x}$ and the Wood-Saxon potential $w(x) = 1/(1+e^{x-1}).$

\section*{Conclusions}

We have reviewed the concept of a resonance in potential scattering in
quantum mechanics in terms of  the Wigner  criterion that not only must the
phase shift pass through $\pi /2$ but there must also be a positive time
delay in the scattering process. In non-relativistic scattering from a
monotone spherically-symmetric potential well that implies that s-wave
resonances do not exist, but that p-wave and higher angular momentum
resonances do exist. The zero energy solutions of the Schr\"{o}dinger
equation provide a starting point for the study of resonances for scattering
off a well: resonances are obtained by slightly weakening the potential
needed to obtain a zero energy solution. 

In the case of relativistic potential scattering using the Dirac equation,
the zero energy solutions are replaced by zero momentum or critical
solutions at $E=m$ and $E=-m$.   Now  resonances are obtained by slightly
weakening the critical potential well giving $E=m$ or slightly strengthening
the supercritical well giving $E=-m.$ In the case where a critical solution
at $E=m$ exists for a potential well, slightly weakening the potential will
give rise to  resonances in s-waves as well as in higher waves. Slightly
increasing the strength of the the supercritical well will give rise to
resonances in scattering from a potential barrier using the crossing theorem,
which relates positive energy solutions of the Dirac equation for a
potential barrier to negative energy solutions for a potential well. We thus
find that resonances exist for scattering from barriers in s-waves and higher
waves. In particular crossing relates s-wave solutions for a well to p-wave
solutions for a barrier and vice versa.

\section*{Acknowledgments}

Partial financial support of this work under grant no. GP3438 from the Natural Sciences and Engineering Research Council of Canada is gratefully acknowledged by one of us (RLH).

\end{document}